\documentclass[traditabstract]{aa}
\usepackage{txfonts,textcomp,natbib,amssymb,longtable,lscape}
\usepackage[dvips]{graphicx}
\bibpunct{(}{)}{;}{a}{}{,} % to follow the A&A style
\newcommand{\Sauron}{\texttt{SAURON}}
\newcommand{\Oasis}{\texttt{OASIS}}
\newcommand{\XSauron}{\texttt{XSauron}}
\newcommand{\XOasis}{\texttt{XOasis}}
\newcommand{\nodata}{...}

\begin{document}

\title{How does star formation proceed in the circumnuclear starburst ring of NGC\,6951?}

%\subtitle{As determined in NGC\,6951}
\author{T.\,P.\,R. van der Laan \inst{1}
\and E. Schinnerer \inst{1}
\and E. Emsellem \inst{2}
\and L.\,K. Hunt \inst{3}
\and R.\,M. McDermid \inst{4}
\and G. Liu \inst{5}
}

\institute{Max-Planck-Institut f\"ur Astronomie, K\"onigstuhl 17, 69117 Heidelberg, Germany; \\ \textit{vanderlaan@iram.fr}
\and European Southern Observatory, Karl-Schwarzschild-Str 2, 85748 Garching, Germany
\and INAF-Osservatorio Astrofisico di Arcetri, Largo E. Fermi 5, 50125 Firenze, Italy
\and Gemini Observatory, Northern Operations center, 670 N. A'ohoku Place, Hilo, HI 96720, USA
\and Center for Astrophysical Sciences, Johns Hopkins University, 3400 North Charles Street, Baltimore, MD 21218, USA
}

\date{Received 24 August, 2012 / Accepted 9 January, 2013}

\abstract{Gas inflowing along stellar bars is often stalled at the location of circumnuclear rings, that form an effective reservoir for massive star formation and thus shape the central regions of galaxies. However, how exactly star formation is proceeding within these circumnuclear starburst rings is subject of debate. Two main scenarios for this process have been put forward: In the first the onset of star formation is regulated by the total amount of gas present in the ring with star forming starting once a mass threshold has reached in a `random' position within the ring like `popcorn'. In the second star formation preferentially takes place near the locations where the gas enters the ring. This scenario has been dubbed `pearls-on-a-string'. Here we combine new optical IFU data covering the full stellar bar with existing multi-wavelength data to study in detail the 580\,pc radius circumnuclear starburst ring in the nearby spiral galaxy NGC\,6951. Using \textit{HST} archival data 
together with \Sauron\, and \Oasis\, IFU data, we derive the ages and stellar masses of star clusters as well as the total stellar content of the central region. Adding information on the molecular gas distribution, stellar and gaseous dynamics and extinction, we find that the circumnuclear ring in NGC\,6951 is $\sim$1-1.5\,Gyr old and has been forming stars for most of that time. We see evidence for preferred sites of star formation within the ring, consistent with the `pearls-on-a-string' scenario, when focusing on the youngest stellar populations. Due to the ring's longevity this signature is washed out when older stellar populations are included in the analysis.}
\keywords{Galaxies: individual: NGC\,6951 - Galaxies: ISM - Galaxies: stellar content}

\titlerunning{Star formation in the starburst ring of NGC\,6951}
\maketitle

\section{Introduction}
Secular evolution comprises all agents of change in a galaxy that are independent of the external environment of the galaxy. Secular evolution, unlike hierarchical clustering and mergers, takes place over long timescales, near-equal to the lifetime of galaxies, and is important for a  galaxy's evolution in the current epoch. A large part of secular evolution takes place due to asymmetries in the gravitational potential of a galaxy. All asymmetries assist in the inward and outward motion of gas across parts of the disk. One of the asymmetries that acts the fastest and most severely is a large scale bar: an elongated concentration of stars and/or gas centered on the nucleus, usually of several kpc size. Depending on the strength of the bar, the sound speed in the gas, and the central mass concentration, a circumnuclear ring will form near the inner Lindblad resonance at a radius defined by the bar \citep[e.g.][]{Athanassoula1992a,Regan2003,2008ApJS..174..337M, 2012ApJ...747...60K}.

A ring directly hinders further inflow towards the nucleus, i.e. nuclear feeding \citep[e.g.][]{Maciejewski2004II,Garcia2005}. A circumnuclear ring is in effect the gas reservoir for the gas moving inward. Determination of the life times of circumnuclear rings is therefore of paramount importance to understand the influence they have on the evolution of their host galaxies. The high gas densities reached in circumnuclear rings naturally lead to star formation. The ages of the stellar population(s) in circumnuclear rings may be a good indicator of the overall age of the ring, and thus the stellar bar itself. A caveat naturally follows; how does star formation proceed in circumnuclear rings? Is there a time scale that can be related to the onset of star formation in the ring (and that should be added to determine the ring's age)?

Much theoretical and modeling work has been done on the formation of circumnuclear rings \citep[e.g. ][]{Athanassoula1992b,Englmaier2000,Regan2003,Maciejewski2004II}. Circumnuclear gas rings are closely related to the more generic nuclear spirals; asymmetries in the gravitational potential induce a wave perturbation in the gas. All gas has angular momentum and it needs to lose this momentum before it can move towards the center. This loss happens in the dust lanes and nuclear spirals.They are the locations where gas is shocked and loses angular momentum, allowing gas to migrate to smaller orbits. 

As has been shown by \citet{Athanassoula1992a,Regan2003} and others, the existence of a circumnuclear ring inside a large scale bar depends, among other factors, on the orbit families that make up the circumnuclear region. $x_{1}$ orbits are elongated along the asymmetric potential and are the main orbits that make up the large scale bar. $x_{2}$ orbits on the other hand, are oriented perpendicular to the bar. Gas orbits slowly turn from one to the other in a bar potential. That $x_{2}$ orbits are again symmetric to the large scale stellar bar potential, is important, because there should be no {\it net} gravitational torques over the orbits (i.e. no further shocks in the gas). $x_2$ orbits are therefore able to serve as repositories for gas. If there is a significant fraction of $x_{2}$ orbits present in the circumnuclear region, gas can settle on them and build up a circumnuclear gas ring.

The connection between dust lanes, nuclear spirals and circumnuclear rings is very clear observationally. The dust/gas lanes of the bar do not extend to the nucleus in the case of a ring and actually end at the ring radius. Two points can often be seen in the gaseous ring, offset by 180\degr\,, where the lanes connect to the ring. They are called `contact points', following \citet{Boker2008}. So, after being shocked and having lost most angular momentum, infalling gas enters the circumnuclear ring at the contact points and will stay on the $x_2$ orbits. The immediate result of this situation is that the gas density in the ring will increase with continued inflow, with star formation as the natural consequence.

The most straight-forward star formation model for circumnuclear rings has been proposed by \citet{Elmegreen1994}. The gas density will build up until it reaches a critical density for star formation. The entire ring will become unstable and star formation will proceed throughout the ring. The time scale of the onset of star formation is then directly related to the inflow rate onto the ring. This scenario was called `popcorn' by \citet{Boker2008}, who themselves proposed another scenario.

This second scenario proposes an azimuthal age gradient for the recently formed stars, starting at the contact points. Since gas enters the ring at two distinct positions, it stands to reason that the gas density in the ring is higher at the contact points (termed `overdensity regions', or `ODR' by the authors) than elsewhere in the ring. The critical density for star formation will be reached first at (or close to) those positions. By overall rotation of the system these newly formed star clusters will be azimuthally distributed in the ring, while star formation continues at the contact points, like `pearls-on-a-string'.

NGC\,6951 is a good candidate to investigate which of these two scenarios holds true and estimate the life time of the circumnuclear ring. NGC\,6951 is a nearby \citep[24.1\,Mpc,][]{Tully} SAB(rs)bc galaxy. It has a circumnuclear ring at a radius of 5\arcsec\ or 580\,pc. It has a low inclination \citep[46.2\degr\,,][]{Haan2009}, which provides a clear view of the entire ring. Its molecular gas dynamics in and around the ring have already been studied in detail by some of the authors \citep{Tessel}, as well as by others \citep{Kohno1999, Krips2007, Storchi2007}. The gravitational torques acting on the cold molecular gas were also analyzed \citep{Haan2009,Tessel}. \citet{2008ApJS..174..337M} have investigated the H$\alpha$ regions in the ring, and found young stellar ages ($<$10\,Myr) in the northern segment of the ring and older ages ($\sim10\,Myr$ for an instantaneous, or $\sim100\,Myr$ for a continuous star formation scenario) in the rest of the circumnuclear ring. No other determination of the ages of the stellar populations of the ring has been published to date.

New observations and archival data retrieved for this study will be presented in \S2. In \S3 the known relevant properties of the ring are summarized. In \S4 the extinction in the ring is derived, before detailing the detection of star clusters in \S5. \S6 contains the determination of the ages of the star clusters and determination of the ages of the overall stellar content based on \Sauron\  data. A discussion with respect to star formation scenarios and ring life time is given in \S7. A summary follows in \S8.

\section{Observations and Archival Data}
\begin{figure}
\resizebox{\hsize}{!}{\includegraphics{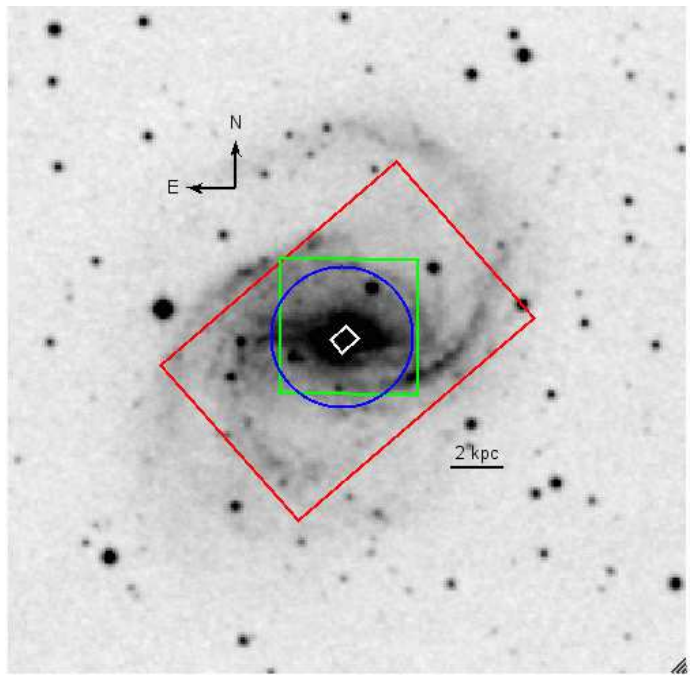}}
\caption{Overlay of the Field-of-View of various datasets presented in this paper. From small to large: \Oasis\ (white), {\it HST} (green), PdBI (blue), \Sauron\ (red). The underlying (B band) image was extracted from DSS.}\label{fig:FoV}
\end{figure}

\begin{table*}
\begin{minipage}{\columnwidth}
\centering
\caption{Characteristics of the \Sauron\ and \Oasis\ datasets.}\label{table:oasis-sauron}

\begin{tabular}{c c c c c c c c}
\hline\hline
IFU & Mode & \# Expo & Exp. Time & Spectral Sampling & Spatial Sampling & Expo FoV & Total FoV \\
\hline
\Oasis\ & MR\,516 & 5 & 30\,min & 2.15\,\AA  & 0.2\arcsec & 10.3\arcsec\, $\times$ 7.4\arcsec & \\ 
\Oasis\ & MR\,661 & 4 & 30\,min  & 2.14\,\AA  & 0.2\arcsec & 10.3\arcsec\, $\times$ 7.4\arcsec & \\
\Sauron\ & LR &  26 & 30\,min & 1.15\,\AA  & 0.8\arcsec & 42\arcsec\, $\times$ 33\arcsec & 135\arcsec\, $\times$ 90\arcsec \\
\hline
\end{tabular}
\end{minipage}
\end{table*} 

\subsection{\Oasis\ data}
The \Oasis\ Integral-Field Unit \citep{2004SPIE.5492..822M} is installed at the William Herschel Telescope (WHT) in La Palma (Canary Islands). Two individual \Oasis\ configurations were used, namely the MR\,516 and MR\,661, which, respectively, cover the spectral range around the H$\beta$, Mg lines ($[4760-5560\AA]$) and around H$\alpha$, [NII], [SII] lines ($[6210-7010\AA]$) at a resolution of $R\sim 1500$, and a sampling of about 2.15\AA\ per pixel (see the \Oasis/WHT page for further details). For both configurations, the 22\,mm enlarger was used, providing a field of view of approximately $10\farcs3 \times 7\farcs4$ (see Fig. \ref{fig:FoV}) with original spaxels of $0\farcs26$.

NGC\,6951 was targeted during a run in July 2006, and we obtained five and four 30\,min exposures, with the MR\,516 and MR\,661 configurations, respectively, centered on the galaxy nucleus. Slight dithering between consecutive exposures was applied to allow for a better rejection of bad pixels or faulty spectra. Details are provided in Table~\ref{table:oasis-sauron}.

All the \Oasis\ data were reduced using the \XOasis\ data reduction software\footnote{http://www-obs.univ-lyon1.fr/labo/oasis/download/}, developed by CRAL (Arlette P\'econtal and coll.), and adapted to the above-mentioned setups. This requires the creation of a spectral mask, following a detailed model of the optical path for the instrument, which allows an accurate optimal extraction of the signal from the raw exposures. The detailed reduction pipeline includes CCD corrections (bias, fringing, removal of the overscans), wavelength calibration, flat-fielding (using both lamp calibrations and sky exposures), and flux calibration. Each individual exposure is then truncated to a common wavelength domain (from 4770\AA\ to 5530\AA\ for MR\,516, and from 6620\AA\, to 6990\AA\ for MR\,661). We then reconstruct images by fully integrating over the spectral direction, and recenter each individual data cube so that the center of NGC\,6951 corresponds to $(0,0)$. The individual exposures are then merged, taking into account the effect of differential refraction, by re-projecting the cubes onto a common spatial grid centered at $(0,0)$ with spaxels of $0\farcs2\times0\farcs2$. The final merged cubes have a spectral sampling of 2.15~\AA\ for the MR\,516 configuration, and 2.14~\AA\ for the MR\,661.

\subsection{\Sauron\ data}
NGC\,6951 was observed with \Sauron\ at the WHT \citep{Bacon+01} during a run in August 2003. Individual exposures cover a field of view of about $41\arcsec \times 33\arcsec$, with spaxels of $0\farcs94$, a spectral resolution of 4.2~\AA\ and spectral sampling of 1.15~\AA\, over the range 4780~\AA\ to 5350~\AA\,. The spectral coverage of the \Sauron\ datacubes overlaps with the \Oasis\ MR\,516, though it is significantly shorter in wavelength coverage. The coarser spatial sampling allows for a much larger spatial coverage of the disk of NGC\,6951 than with \Oasis, see also Table \ref{table:oasis-sauron}. We obtained a total of twenty-six 30\,min exposures, the goal being to cover as much of the stellar bar as possible. The total time spent on each pointing was adopted according to the average surface brightness, and included some slight dithering for each pointing to prevent systematics. The final area covered by the complete mosaic is $~135\arcsec\times90\arcsec$ (see Fig. \ref{fig:FoV}). The data from the single, central exposure has been presented previously by \citet{2007MNRAS.379.1249D}.

The data reduction of all \Sauron\ exposures was performed using a pipeline based on \XSauron\, which is similar to the \XOasis\ version, but adapted to the different optics and setup. The reduction steps are thus very similar to the ones described in the previous Section for \Oasis\,. The biggest differences lie in the characteristics of the spectral mask, and the existence, for \Sauron\,, of about 150 dedicated sky lenses which allow for an accurate sky subtraction for each exposure.

The fully calibrated individual exposures were then recentered (using a direct image of the galaxy as a reference) and merged, by projecting each data cube on a common grid with $0\farcs8\times0\farcs8$ spaxels. To increase the S/N, the spaxels were Vonoroi binned to a S/N of 40.

\subsection{Spectral fitting of the IFU datasets}\label{sec:gandalf}
We applied standard routines to extract the kinematics: the stellar kinematics (velocity, velocity dispersion and Gauss-Hermite $h_3$ and $h_4$ values) with pPXF \citep{CappEms04, Emsellem+04} using a set of stellar templates extracted from the MILES library\footnote{http://miles.iac.es} \citep{Sanchez+06,2011A&A...532A..95F} and the IDL implementation available from Michele Cappellari's web page\footnote{http://www-astro.physics.ox.ac.uk/$\sim$mxc/idl/}, and the gas kinematics via a version of GANDALF \citep[kindly adapted for our purpose by Marc Sarzi,][]{Sarzi+06}. Fitted emission lines include H$\beta$, the doublets [OIII] ($\lambda$ (4959,5007) $\AA$), [NI] ($\lambda$ (5200,5202) $\AA$), in both \Sauron\,, and \Oasis\ MR\,516, and [OI] ($\lambda$ (6300,6364) $\AA$), H$\alpha$ ($\lambda$ 6562.8 $\AA$), and the doublets [NII] ($\lambda$ (6548,6583) $\AA$) and [SII] ($\lambda$ (6717,6731) $\AA$) for the \Oasis\ MR\,661 data cube.  The kinematics of the emission lines was fixed to the [OIII] doublet (\Sauron\,) and [OI], [NII] and [SII] kinematics were fixed together in \Oasis\ MR\,661. The H$\alpha$ emission was strong enough to be fit independently. We emphasize that H$\beta$ and H$\alpha$ are present both in emission and absorption, however, the combined use of pPXF and GANDALF does take this into account.

\begin{figure*}
\centering
\includegraphics[width=17cm]{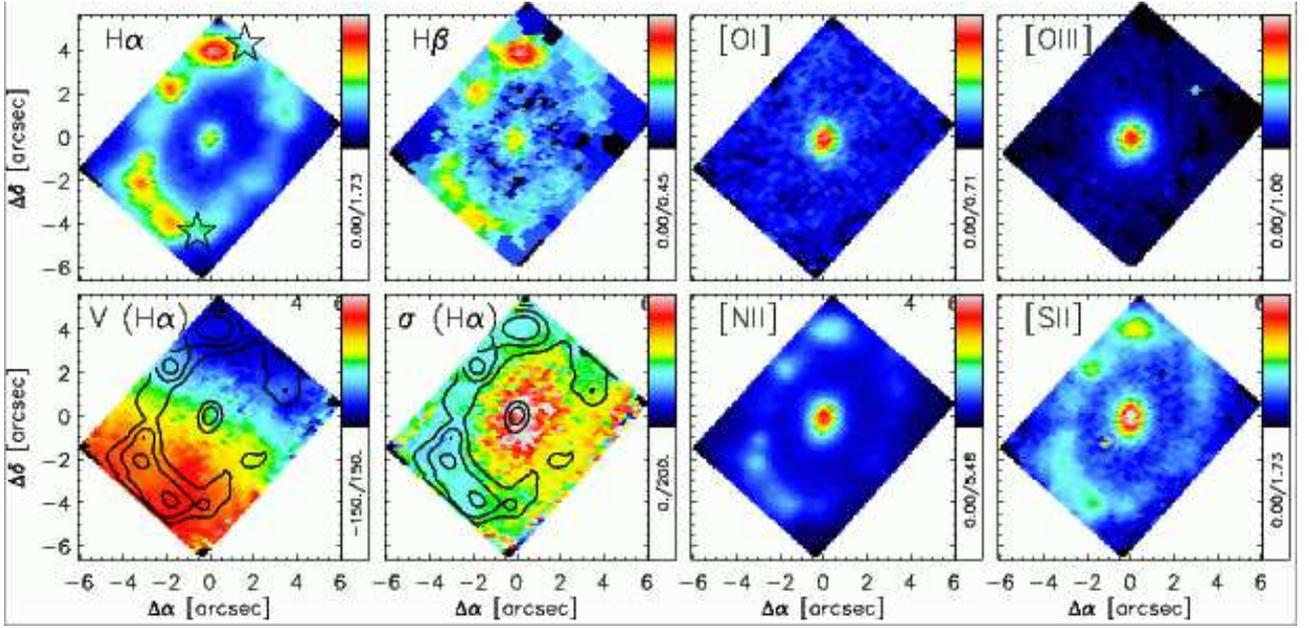}
\caption[NGC\,6951 \Oasis\, maps]{{\it Top panels and bottom right:} \Oasis\ intensity maps of the emission lines H$\alpha$, H$\beta$, [OI], [OIII], [NII] and [SII].  The color scale in each panel indicates the square of the intensity, the range of which is given below the color bar. This scaling was chosen to emphasize faint structures in the maps. The `contact points', discussed in the text, are indicated with a star in the H$\alpha$ panel. {\it Bottom left panels:} the velocity and velocity dispersion map of H$\alpha$. The flux contours of the integrated H$\alpha$ emission are over plotted for emphasis. The velocity range in the bottom-left panel is [-150, 150] km s$^{-1}$, with an assumed systemic velocity of NGC\,6951 of 1425 km/s. The velocity dispersion range in the adjacent panel is [0, 200] km/s. The orientation of all maps is north up, east left. All positions are relative to the nucleus.}\label{fig:OASIS}
\end{figure*}

The resulting emission line maps from the spectral fitting for \Oasis\ are shown in Fig. \ref{fig:OASIS}. It can be seen that the \Oasis\ FoV just covers the circumnuclear ring. The H$\alpha$ and H$\beta$ emission is mostly concentrated in the ring, while the [OI] and [OIII] emission is limited to the nucleus. The [NII] and [SII] emitting gas is present in both the nucleus and ring. There are bright hot-spots in the ring that are seen in H$\alpha$, H$\beta$, [NII] and [SII]. The velocity field and dispersion are only shown for H$\alpha$, but are very similar for the other emission lines. The increase in the velocity dispersion in the western half of the ring was also seen in the \Sauron\ observations presented by \citet{2007MNRAS.379.1249D} and coincides with the lowest intensities of H$\alpha$ and H$\beta$ line emission in the ring.

\begin{figure*}
\centering
\includegraphics[width=17cm]{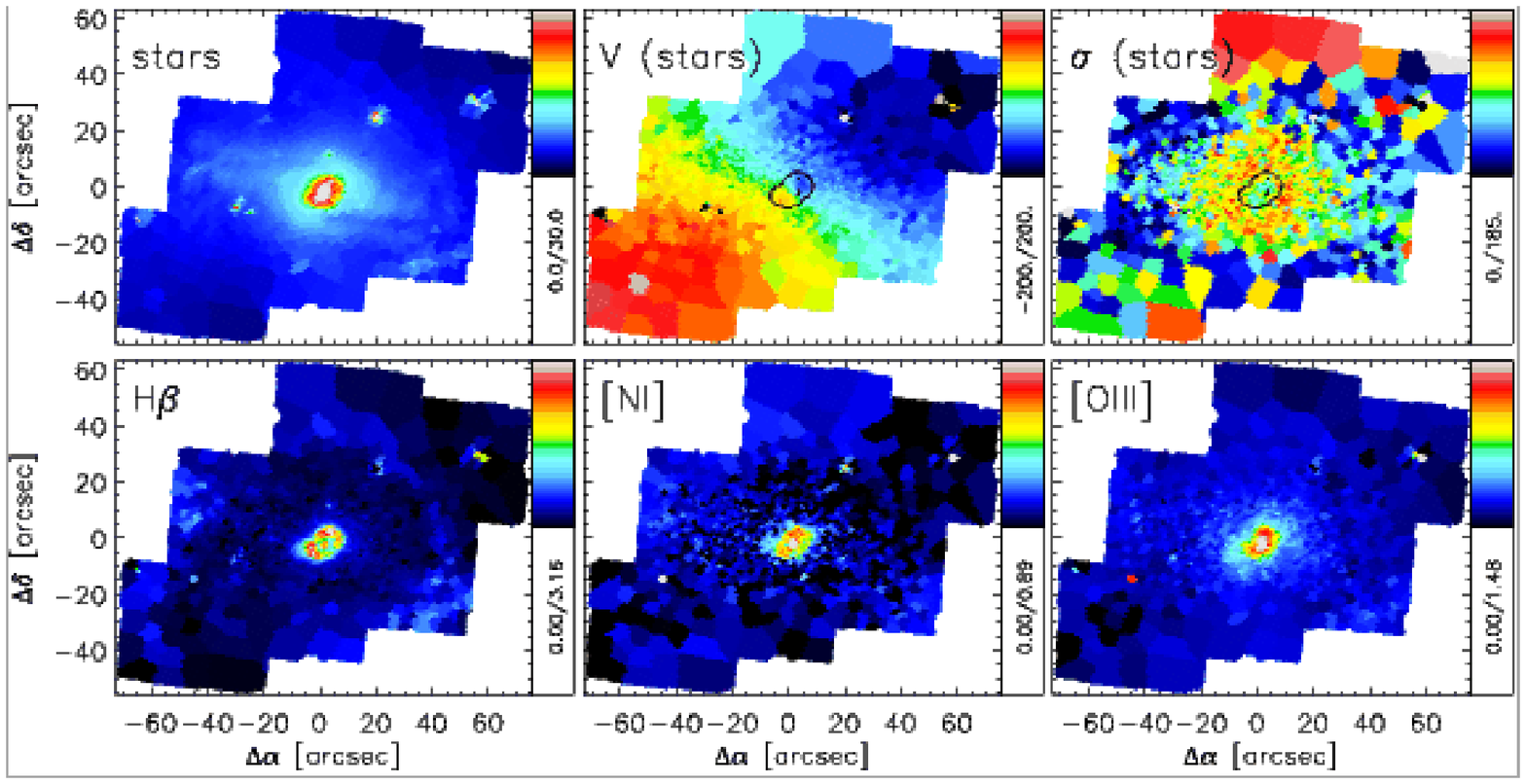}
\caption[NGC\,6951 \Sauron\, maps]{\Sauron\ intensity maps of the stellar light ({\it top left}), H$\beta$ ({\it bottom left}), [NI] ({\it bottom middle}) and [OIII] ({\it bottom right}) emission lines, as well as the stellar velocity field ({\it top middle}) and velocity dispersion map ({\it top right}). The color scale in each panel indicates the square of the intensity, or the velocity, the range of which is given below the color bar. The orientation of the maps is north up, east left. All positions are relative to the nucleus. To guide the eye, the contour corresponding to the central H$\beta$ emission has been over-plotted onto the velocity and velocity dispersion plots. The velocity range in the top-middle panel is [-200, 200] km s$^{-1}$, the assumed systemic velocity of NGC\,6951 is 1425 km/s. The velocity dispersion range in the top-right panel is [0, 185] km/s.}\label{fig:SAURON}
\end{figure*}

The spectral fitting results from the \Sauron\ mosaic are presented in Fig. \ref{fig:SAURON}. In the top-left panel of Fig. \ref{fig:SAURON} we show the total intensity over the FoV of the mosaic. The peak intensity of the central region has been saturated to show the details of the entire stellar bar region. The stellar bar is orientated approximately east-west on the sky. The few high-intensity peaks are foreground stars. The stellar velocity field and dispersion are also shown in the top row. 

In the bottom row of Fig. \ref{fig:SAURON} the H$\beta$, [NI] and [OIII] emission line maps are presented. It is clear that their emission is concentrated in the circumnuclear region. In the case of H$\beta$, a keen eye can make out the ring shape of the emission in this figure (exactly as was seen in the \Oasis\ map). 

\begin{table*}
\begin{minipage}{1.5\columnwidth}
\caption{\textit{HST} archival data \label{tab:sampledetails}}
\begin{tabular}{l l l c c l c}
\hline\hline
Filter & Program ID & Exposure [s] & Obs. Date & PHOTFLAM & Instrument & Scale [\arcsec/pixel] \\
\hline
%F190N & Pa$\alpha$ cont. & 1343.763 & 13-20-2007 & & NICMOS NIC3 & 0.10''\\
F187N$^{*}$ & 11080 & 1151.767 & 13-02-2007 & 3.6837500$\times$10$^{-18}$ & NICMOS NIC3 & 0.10\\
F160W & 7331 & 255.923 & 16-12-1997 & 2.3600094$\times$10$^{-19}$ & NICMOS NIC2 & 0.076\\
F110W & 7331 & 255.923 & 16-12-1997 & 4.3320170$\times$10$^{-19}$ & NICMOS NIC2 & 0.076\\
F547M & 5419 & 300 & 16-07-1994 & 7.691168$\times$10$^{-18}$ & WFPC2/PC & 0.046\\
F606W & 8597 & 160 & 18-11-2000 & 1.899711$\times$10$^{-18}$ & WFPC2/PC & 0.046\\
F658N & 9788 & 700 & 24-10-2003 & 1.9597158$\times$10$^{-18}$ & ACS WFC1 & 0.05\\
F814W & 9788 & 120 & 24-10-2003 & 6.9255574$\times$10$^{-20}$ & ACS WFC1 & 0.05\\
F330W & 9379 & 1200 & 17-08-2002 & 2.2688470$\times$10$^{-18}$ & ACS HRC-FIX & 0.025\\% ergs/cm2/Ang/electron
\hline
\end{tabular} \\

\textbf{Notes:} List of the observations extracted from the \textit{HST} archive. $^{*}$The Pa$\alpha$ image has been corrected for continuum flux with a F190N-filter image taken at the same time.
\end{minipage}
\end{table*} 

\subsection{\textit{HST} images}
We retrieved F547M, F606W, F658N, F814W and F330W images of NGC\,6951 from the MAST \textit{HST} archive. We extracted them from the archive calibrated with the "best" reference files. We also utilize NICMOS F110W and F160W images from the \textit{HST} archive already used in previous work \citep[][original: \citet{2004ApJ...616..707H}]{Tessel}. Finally, we obtained a Pa$\alpha$ image, which was observed during the same campaign and reduced using the same approach as for NGC\,1097 in \citet{2011ApJ...736..129H}. The full sample and detailed data reduction for the Pa$\alpha$ image are presented in Liu et al. (in prep.). A list of all images used is given in Table \ref{tab:sampledetails}. 

The H$\alpha$ image was carefully continuum subtracted with the F606W image, as was the Pa$\alpha$ (F190N) image using the F187N image. The F658N filter passband includes emission from the [NII] double line emission. From spectra presented in \citet{Storchi2007} the ratio [NII]6583\,\AA\,/H$\alpha$ is known to be $\sim$\,0.3 throughout the ring, implying a 37\% contribution to the observed flux from [NII]6548\,\AA\, and [NII]6583\,\AA\, in the {\it HST} F658N filter for a V$_{sys}$ of 1425 km/s, for which we correct.

\begin{figure}%[htpb]
\resizebox{\hsize}{!}{\includegraphics{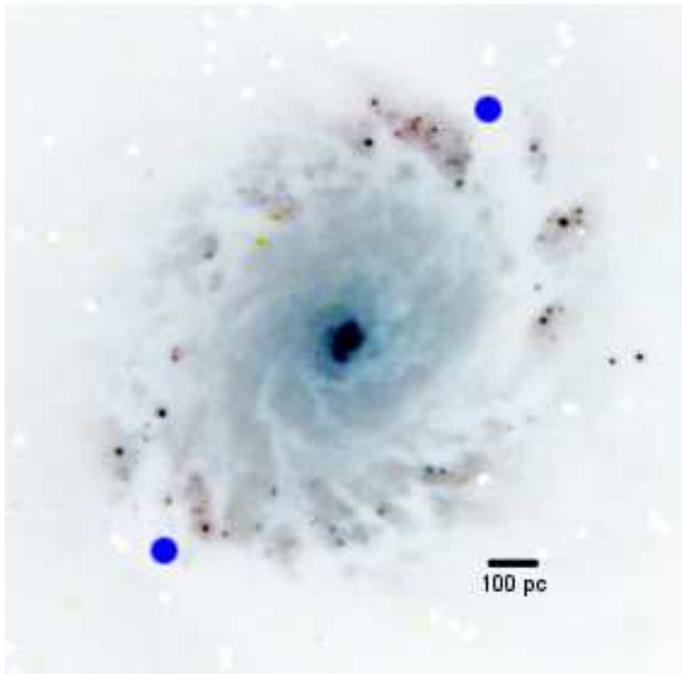}}
\caption{False color image of the circumnuclear ring in NGC\,6951, based on the F814W (red), F606W (green) and F547M (blue) broad band filters. The image ($\sim14\arcsec\, \times \sim14\arcsec$) is orientated with north pointing up and east to the left. The center has been saturated for better visibility of the ring, and bad pixels have been removed. The `contact points' of the ring are indicated with blue circles.} 
\label{fig:prettycolor}
\end{figure}

\section{NGC\,6951's circumnuclear ring so far...}
\begin{table}
\begin{minipage}{\columnwidth}
\caption{Properties of NGC\,6951}\label{table:theovals}
\begin{tabular}{l l l}
\hline\hline
Parameter & Value & Reference \\
\hline
Type & SAB(rs)bc & (1)\\
Nuclear Activity & LINER/Seyfert 2 & (2) \\
\multicolumn{2}{l}{Dynamical Center (Locus Radio Continuum)}&\\
\,\,\,\,RA (J2000) & 20$^h$37$^m$14.123$^s$ & (3,4) \\
\,\,\,\,Dec (J2000) & 66\degr06\arcmin20.09\arcsec & (3,4)\\
Inclination Angle & 46.2\degr & (5)\\
Position Angle & 138\degr & (5)\\
Adopted Distance & 24.1 Mpc & (6)\\
Stellar bar PA & 84\degr & (7)\\
Mass mol. gas (r$<$3\,kpc) & 1.6 $\times 10^9$ M$_{\odot}$ & (8) \\
Mol. gas inflow rate onto ring & 2.0M$_{\odot}$/yr & (8) \\
Large scale bar pattern speed & 28.9$\pm$1.2 km/s/kpc & (9)\\
V$_{sys}$ & 1425 km/s & (5) \\
Period pattern speed & 210\,Myr & (9) \\
1 rotation for a ring orbit & 24\,Myr & (9) \\
\hline
\end{tabular}\\

\textbf{References:} (1) \citet{deVaucouleurs}, (2) \citet{Perez2000}, (3) LEDA, \citet{Krips2007}, (4) \citet{Saikia2002}, (5) \citet{Haan2009}, (6) \citet{Tully}, (7) \citet{1997ApJ...482L.135M}, (8) \citet{Tessel}, (9) this work
\end{minipage}
\end{table}

Before the start with the analysis of the stellar properties of NGC\,6951's circumnuclear ring, a brief summary of the gaseous properties of the circumnuclear region that are known to date is presented. For clarity later on, all numbers pertinent to the discussion later are also listed in Table \ref{table:theovals}. The large scale stellar bar in NGC\,6951 has a PA of 84\degr\, on sky \citep{1997ApJ...482L.135M}. Its semi-major axis is 3.0\,kpc. The dust lanes of the bar are thus orientated almost horizontally on the sky. The contact points in the ring are in the north and south of the ring, see also Figs. \ref{fig:OASIS} and \ref{fig:prettycolor}. 

$^{12}$CO(1-0) and $^{12}$CO(2-1) observations, taken with the IRAM PdBI and 30m telescope, of the circumnuclear region of NGC\,6951 are presented by \citet{Tessel}. The resolution of the CO(1-0) observations is 3.11\arcsec\, by 2.59\arcsec\,, the CO(2-1) observations are 1.72\arcsec\, by 1.56\arcsec. The FoV of the CO(1-0) observations is shown in Fig. \ref{fig:FoV}, the FoV of the CO(2-1) data is about half that size. Both completely cover the ring. Their r.m.s. noise corresponds to M$_{H_{2}} \sim 10^5$ M$_{\odot}$/beam. We refer to \citet{Tessel} for further details on the data reduction. A total H$_2$ molecular gas mass of 2.2 $\times 10^9$ M$_{\odot}$ was derived from the CO(1-0) observations, assuming a Galactic conversion factor and corrected for helium abundance. An HCN(1-0) intensity map obtained with PdBI was presented by \citet{Krips2007}. A map with slightly different $uv$-weighting from that dataset was kindly provided by M. Krips (priv. comm.), with a beam size of 1.89\arcsec by 1.83\arcsec (PA 78.7\degr). 

The key points for the molecular gas morphology are prominent gas spiral arms, that are co-spatial with the dust lanes along the large-scale stellar bar and end at the radius of the circumnuclear ring. The CO(2-1) observations, which have a higher spatial resolution, show that the spiral arms continue into a ring. The HCN(1-0) traces higher density molecular gas than CO, and the HCN intensity distribution shows a similar morphology to the CO(2-1) molecular gas. However, the intensity maxima are at slightly different positions in the ring. \citet{Tessel} derived, from modeling the orbits in the circumnuclear region, that the inflow rate of gas onto the ring is $\sim$2.0 M$_{\odot}$ per year. 

The HI distribution, presented in \citet{Haan2009}, shows that the circumnuclear region of NGC\,6951 is HI-deficient. The neutral ISM in the circumnuclear region is thus molecular gas dominated. Given an estimate for the dynamical mass at 1\,kpc, (M$_{dyn} = \frac{r v^2}{G} \approx 8 \times10^{9} M_{\odot}$, with v=200km/s), atomic and molecular gas makes up approximately 25\% of the dynamical mass within this radius.

Two determinations of the large scale bar pattern speed of the stellar bar in NGC\,6951 exist in the literature. \citet{1998AJ....116.2136A} obtained a value of $\Omega_p$ = 30.3 km/s/kpc (when compensating for the difference in adopted distance) based on the rotation curve and location of the corotation radius and \citet{Haan2009} derived a value of $\Omega_p$ = 23 km/s/kpc based on an analysis of dynamic resonances. The \Sauron\ IFU mosaic covers most of the large scale stellar bar of NGC\,6951, and is ideally suited to measure the pattern speed using the model-free Tremaine-Weinberg method \citep{TW1984}. Using the \Sauron\ mosaic gives a value of 28.9$\pm$1.2 km/s/kpc, corrected for inclination (46.2\degr) and at a distance of 24.1\,Mpc. Details will be available in a forthcoming paper.

With a pattern speed of 28.9\,km/s/kpc, the pattern will take 210\,Myr to complete one rotation. The gas and stars rotate with a velocity of $\sim$150\,km/s at the ring radius. They will thus complete a rotation in the ring in 24\,Myr.

\section{Extinction in the ring}\label{sec:extinction}
From the three Hydrogen recombination lines (H$\alpha$, H$\beta$ and Pa$\alpha$) in our data sets we can compute a spatial distribution of the extinction in the ring. While determining the extinction is not directly crucial to the overarching goal in this work, it provides a consistency check on the derived color excesses of the star cluster fitting (Sect. \ref{sec:starclusters}). We use two different ratios to compute the extinction, the H$\alpha$/H$\beta$ ratio (as derived from our \Oasis\ data) and the H$\alpha$/Pa$\alpha$ ratio (as derived from the \textit{HST} data). 

\begin{figure*}
\centering
\begin{tabular}{c c}
\includegraphics{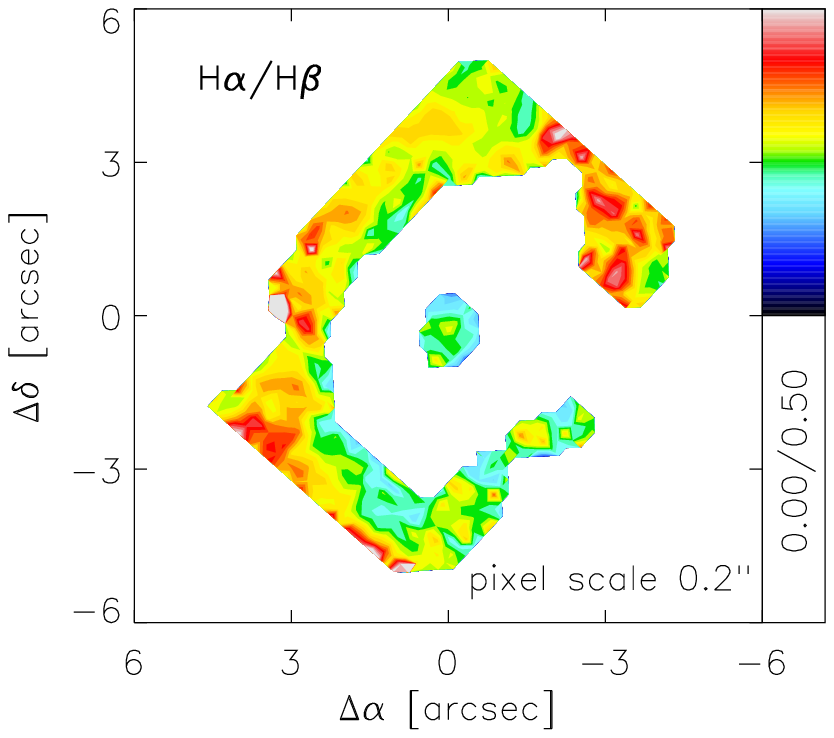}
&
\includegraphics{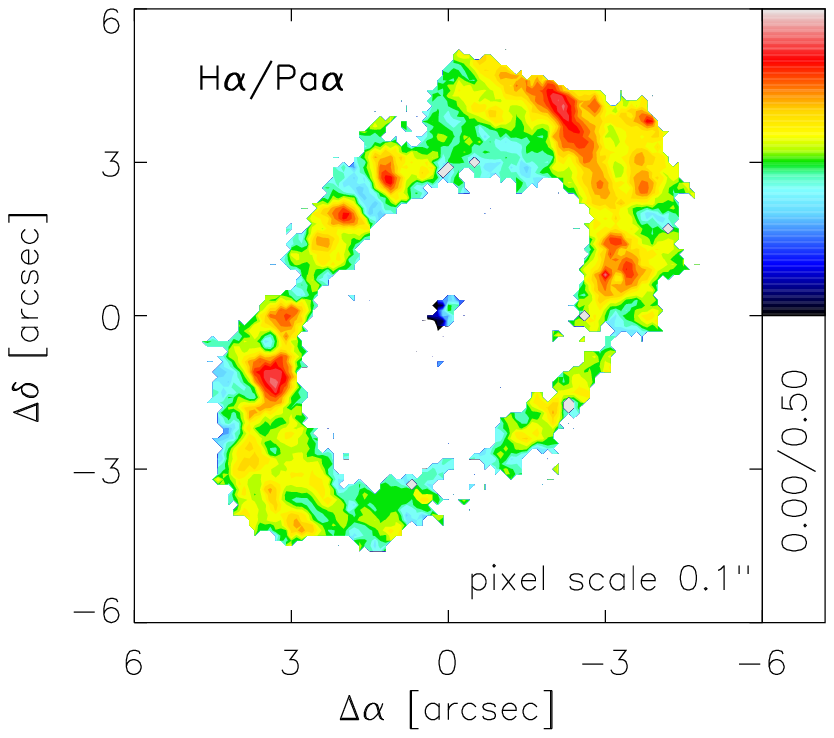}
\\
\end{tabular}
\caption[Color excess E(B-V) from hydrogen emission lines]{{\it Left}: Color excess E(B-V) in the ring as measured from the H$_{\alpha}$/H$_{\beta}$ line ratio using the \Oasis\ line maps. {\it Right}: Color excess in the ring as measured from the H$\alpha$/Pa$\alpha$ line ratio using \textit{HST} images. The pixel scales of each map are given in the lower-right corner of each panel.}
\label{fig:extinction}
\end{figure*}

Both the H$\alpha$ and H$\beta$ maps are obtained during the spectral fitting of the \Oasis\ data cubes. The fitting takes H$\alpha$ and H$\beta$ stellar absorption into account as explained in Sect. \ref{sec:gandalf}. In the case of case B recombination, the ratio H$\alpha$/H$\beta$ will be 2.86 \citep{Osterbrock1989}. Extinction will increase the value. In individual pixels, we find ratios as high as $\sim$4, which is equal to an extinction of $\sim$\,1 magnitude. Inside the ring, the surface brightness of especially H$\beta$ is so low, that no reliable H$\alpha$/H$\beta$ ratio can be computed. Where that is the case, the position has been left blank. In Fig. \ref{fig:extinction} (\textit{right}) the color excess, E(B-V), based on the H$\alpha$/H$\beta$ ratio is plotted. Color excess is a wavelength independent measure of extinction, and is related to extinction via E(B-V) = A$_V$/R$_V$, where R$_V$ gives the total-to-selective extinction, which we assume to be 4.05 \citep[starburst value of][]{Calzetti2000}. 

The FoV of \Oasis\ only barely covers the entire ring. Nevertheless, it is possible to see some trends in the H$\alpha$/H$\beta$ map. The extinction is clearly higher near and `down-stream' from the contact points between ring and bar-induced gas lanes (galaxy rotation is clockwise).

The ratio of the \textit{HST} H$\alpha$ and Pa$\alpha$ observations provides another measure of the extinction. The advantages of these observations are the complete view of the ring and higher spatial resolution (Fig. \ref{fig:extinction}, \textit{left}). Without extinction the expected ratio H$\alpha$/Pa$\alpha$ is about 8.5 \citep[case B recombination]{Hummer1987}. Inside the ring the intensity of both H$\alpha$  and Pa$\alpha$ is too low to obtain reliable ratios. Those positions have been left blank. Again we see the highest extinctions `down-stream' from the contact points. In the western part of the ring we also find `pockets' of higher extinction which can be related to the highest intensity CO and HCN peaks. These places are heavily obscured and dense, and thus are likely the sites of future star formation. Comparison with the H$\alpha$ distribution shows the brightest H$\alpha$ regions to be anti-correlated with extinction.

The H$\alpha$/H$\beta$ based color excess map (\Oasis) and the H$\alpha$/Pa$\alpha$ based color excess map (\textit{HST}) show similar distributions, and have similar values in the range [0.2\,-\,0.5] mag.

\section{Star clusters}\label{sec:starclusters}
Star clusters are a prime way to study the ages of stellar populations; all the stars within a cluster are coeval and clusters are bright and compact, thus easily observable even over large distances. At the distance of NGC\,6951 star clusters are point sources and can easily be separated from the background light in the galaxy. 

\subsection{Identification}
Two independent routines were used for the detection and photometry of the star clusters in the ring of NGC\,6951. These routines are the DAOPHOT package \citep{Stetson1987,1990ASPC....8..289S,1992ASPC...25..297S} in IRAF and SExtractor \citep{1996A&AS..117..393B}. With this approach, the strengths of both methods can be exploited to obtain the most complete star cluster candidate list.

For the detection of the star clusters a composite optical image was created out of 4 individual optical bands; F547M, F606W, F814W and F330W. First, the F606W, F814W and F330W images were aligned with the F547M image using standard procedures in IRAF. For the F814W and F330W bands the images were resampled to the F547M pixel size while conserving the flux. The four resulting images were then normalized and added together.

Applying DAOFIND, 33 star clusters candidates were detected in the composite optical image (within r $\sim$20\arcsec). The star cluster detection in the NIR images was done separately from the composite optical image, as the NIR images have a lower angular resolution. The F110W image was used to detect the NIR star clusters and the resulting positions were applied to the F160W image. Most of the NIR detections have a counterpart in the composite optical image, but several sources were detected without one. This leads to a total of 42 detected star clusters candidates using the DAOPHOT method.

Next, SExtractor was run. The strength of SExtractor is its ability to 'de-blend' crowded regions. As can be seen from the false color image (Fig. \ref{fig:prettycolor}), especially the northern region of the ring is very crowded. For detection and the subsequent photometry of star cluster locations the same composite optical image from the DAOFIND analysis was used. SExtractor automatically determines apertures around the selected source. If the individual images were used independently, the selected apertures between the different wavelengths could vary, which would introduce systematic changes in our photometry. By using the composite optical image as the `detection image'  this is avoided and each source will have the same aperture at the different wavelengths. 

The SExtractor output is filtered by discarding all detections flagged as `bad' by SExtractor itself (e.g. likely to be cosmic rays or one or more pixels that are saturated). The likely candidates are further constrained by location (a rectangular area around the ring of 12\arcsec\, $\times$ 12\arcsec\,) and by relative error in the measured flux ($<$20\%). 43 star cluster candidates are detected. 

In total, 55 different star clusters candidates are detected (see Table \ref{tab:fluxiraf}), about half of which are detected by both DAOPHOT and SExtractor (28 out of 55 or 51\%).

\begin{figure}
\resizebox{\hsize}{!}{\includegraphics{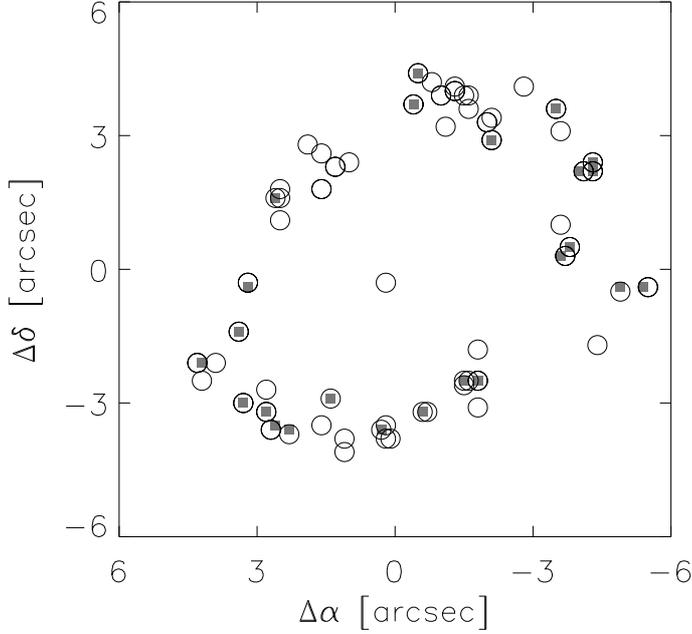}}
\caption[Locations of the candidate star clusters in NGC\,6951]{Locations of the candidate star clusters as detected with IRAF/DAOPHOT and SExtractor (open circles). We have also indicated the locations of the star clusters found by \citet{Barth1995} (grey squares). The axes denote the projected distances in arcseconds from the photometric nucleus, see Table \ref{table:theovals}.}
\label{fig:sourcesiraf}
\end{figure}

\subsection{Properties}
The 55 star cluster candidates can be compared with the work by \citet{Barth1995}, who used a single F547M image (the same one we extracted from the archive). All 24 sources already presented in that work are also included in our list. The additional candidates lie in the northern region and several have lower surface brightnesses. Investigation of the F330W image, which has the finest native pixel scale, shows that some of our detections, particularly in the crowded northern region of the ring, are actually still composed of two or multiple point sources/star clusters.

SExtractor simultaneously detects the star cluster candidates and measures their flux. For the star cluster candidates identified with DAOFIND, their photometric fluxes were obtained using PHOT (also in the IRAF/DAOPHOT package). For the measurement the F330W and F814W images are used at the resolution of the F547M and F606W images. In PHOT the circular apertures for the optical images were chosen to have a radius of 2 pixels (0.09''). This radius contains most of the observed PSF (4 pixels\,$\approx$\,FWHM\,$\sim$\,20\,pc), while minimizing the contamination from other sources (such as other close-by star clusters). The sky subtraction was based on an annulus around each source with radii between 6 and 16 pixels. Flux outliers in this annulus (e.g. other clusters or cosmic-rays, which would over-estimate the sky-levels) were automatically excluded from the sky estimate. For the NIR images an photometry aperture radius of 1.2 pixels was used, which corresponds to a size equal to the optical aperture. The radii of the sky estimate annulus were again 6 and 16 pixels for the NIR images.

The TinyTim package\footnote{http://www.stsci.edu/software/tinytim/tinytim.html} was used to obtain the necessary aperture corrections for each band. The aperture corrections are for the F330W band $1.16\times f_{330}$, F547M band $1.29\times f_{547}$, F606W band $1.31\times f_{606}$, F658N band $1.40\times f_{658}$, F814W band $1.36\times f_{814}$, F110W band $1.38\times f_{110}$ and F160W band $1.59\times f_{160}$.

All measured photometric fluxes are converted into absolute magnitudes, using a distance modulus of 31.9 mag. \citet{Barth1995} used a distance modulus of 31.2, which corresponds to a distance of 17.4\,Mpc. Assuming a distance of 24.1\,Mpc to NGC\,6951, as done in \citet{Tessel} increases the distance modulus to 31.9, or correspondingly increases the magnitudes by -0.7\,mag. The absolute magnitudes are given in Table \ref{tab:fluxiraf}.

The absolute magnitudes from both methods are in good agreement with each other and with \citet{Barth1995}, with at most a 0.5 mag difference in the optical bands and 1\,mag in the NIR bands, mostly due to differences in sky subtraction. These values have not (yet) been corrected for extinction. A measure for the intrinsic extinction is derived during the star cluster fitting.

\section{Ages of the stellar populations in the ring}\label{sect:stellarages}
\begin{figure*}
\includegraphics[width=17cm]{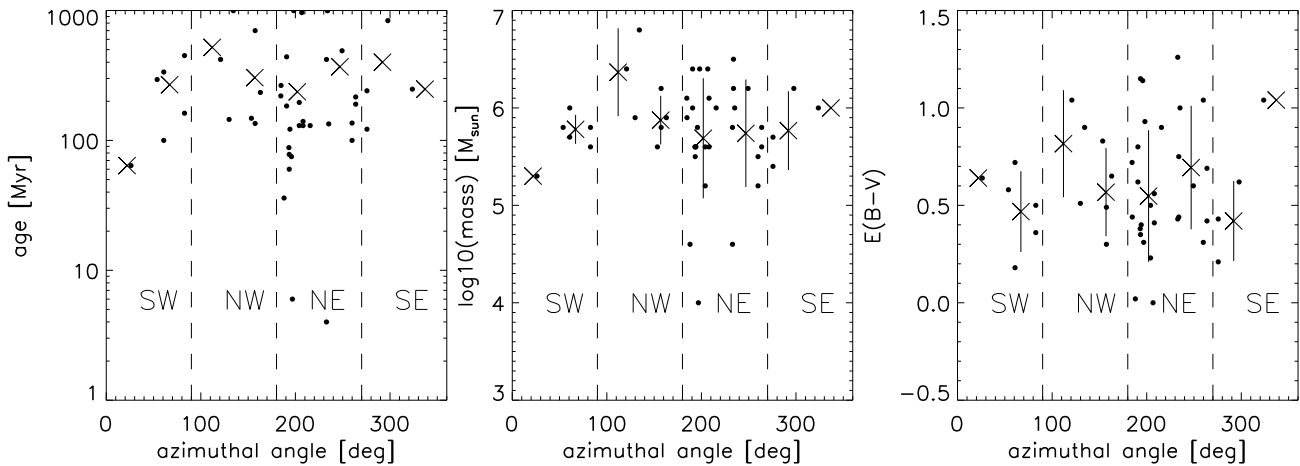}
\caption{Results of the $\chi^2$ fitting. The four segments of the ring are indicated in the plots. From left to right are shown: stellar age, mass and color excess of the cluster. The individual results are averaged every 45\degr (crosses) and the variance is indicated with the error bars. The angle starts at the southern point of the ring and is measured counter-clockwise.}
\label{fig:starclusterresults}
\end{figure*}

Determination the ages of the stars in the ring gives information on the age of the ring in NGC\,6951 itself, since the star formation occurs in the circumnuclear ring. 
\subsection{Starclusters}\label{sect:starclusterages}

The online Starburst99 (v6.0.2) code \citep{1999ApJS..123....3L,2005ApJ...621..695V,2010ApJS..189..309L} was run to obtain model SEDs for star clusters of different ages. We assumed a fixed stellar mass of 10$^{5}$ M$_{\sun}$, an instantaneous star formation event with a Kroupa IMF and Padova AGB solar metallicity stellar tracks. The model SEDs span the UV to the NIR in the time range of 10$^{4}$ to 10$^{9}$ years in steps of 2$\times$10$^{6}$ years. We also include nebular emission in the model fluxes, which adds significant emission for young ages at, predominantly, the shorter wavelengths.

For comparison to our observed data points, these model spectra are weighted with a range of E(B-V) values ([0, 3] in 0.01 steps) and stellar masses ([10$^3$,10$^7$ M$_{\sun}$],  in 0.2 dex steps). Mass scales linearly with luminosity and as such scales to the observed SED, independent of wavelength. A different distance modulus (31.2 instead of 31.9) would have a similar effect. To convert the color excess into an extinction curve the Calzetti extinction law \citep{Calzetti2000} for starburst galaxies is used. The Starburst99 spectra so obtained are convolved with the different \textit{HST} filter throughput curves, and converted to magnitudes. A $\chi^2$-fit is applied to each star cluster set of observed magnitudes. The lowest $\chi^2$ value is selected as the best fit to our data (Table \ref{tab:clusterfitresult}).

Table \ref{tab:clusterfitresult} shows that the best fit ages for the star clusters span young (4\,Myr) to intermediate (300\,Myr) ages, with several older clusters present ($\ga$700\,Myr). The stellar mass range of the star clusters is $\sim$10$^{4.0}$ to $\sim$10$^{6.8}$ M$_{\sun}$ and the E(B-V) color excess varies from 0 to 1.0 magnitudes, with most star clusters having a color excess of $<$0.6. The results of the star cluster fitting are also shown as a function of azimuthal position in the ring in Figure \ref{fig:starclusterresults}. The four quadrants (SW, NW, NE and SE) should be visualized as overlaying compass points on the ring and are chosen with the contact points between gas/dust lanes of the large scale bar and the ring in mind. These points fall just inside the SW and NE quadrants. The rotation sense of the galaxy is clockwise, which corresponds to an decreasing angle. The results are averaged every 45\degr\ to detect any azimuthal trends in age, mass or color excess. 

The star clusters have predominantly intermediate age and are massive, which explains why they are still detectable. From the left panel (age) in Fig. \ref{fig:starclusterresults} it can be seen that the star cluster ages in the ring vary little, although they do drop in the SW and NE quadrant, before rising. It is not clear whether this trend could be a weak age gradient. \citet{2008ApJS..174..337M} find a qualitatively similar age distribution, although we should point out that $\sim100\,Myr$ ages for the H$\alpha$ emission region in that work can only be achieved with a continuous star formation history. The star clusters are assumed to have formed in a single burst. However, the H$\alpha$ regions discussed in \citet{2008ApJS..174..337M} cover a much larger area than the individual star clusters in this work. Thus, both results would be in agreement if some younger stellar population is not or no longer concentrated in star clusters, but is present in the ring. We discuss this scenario further in the next subsection.

The color excess derived from the star cluster fitting (on average 04-0.6\,mag) is somewhat larger than the color excess values obtained from the H$\alpha$/H$\beta$ and H$\alpha$/Pa$\alpha$ ratios presented in Sect. \ref{sec:extinction} (on average 0.4\,mag). However, the hydrogen emission and the star clusters do not come from the same physical locations. The extinction that is found for the star clusters is confined to the star cluster region, and can thus be higher than the more global extinction map derived from the Hydrogen lines.

\subsection{Stellar populations from \Sauron\ IFU}\label{sect:IFUstar}
\begin{figure*}
\includegraphics[width=17cm]{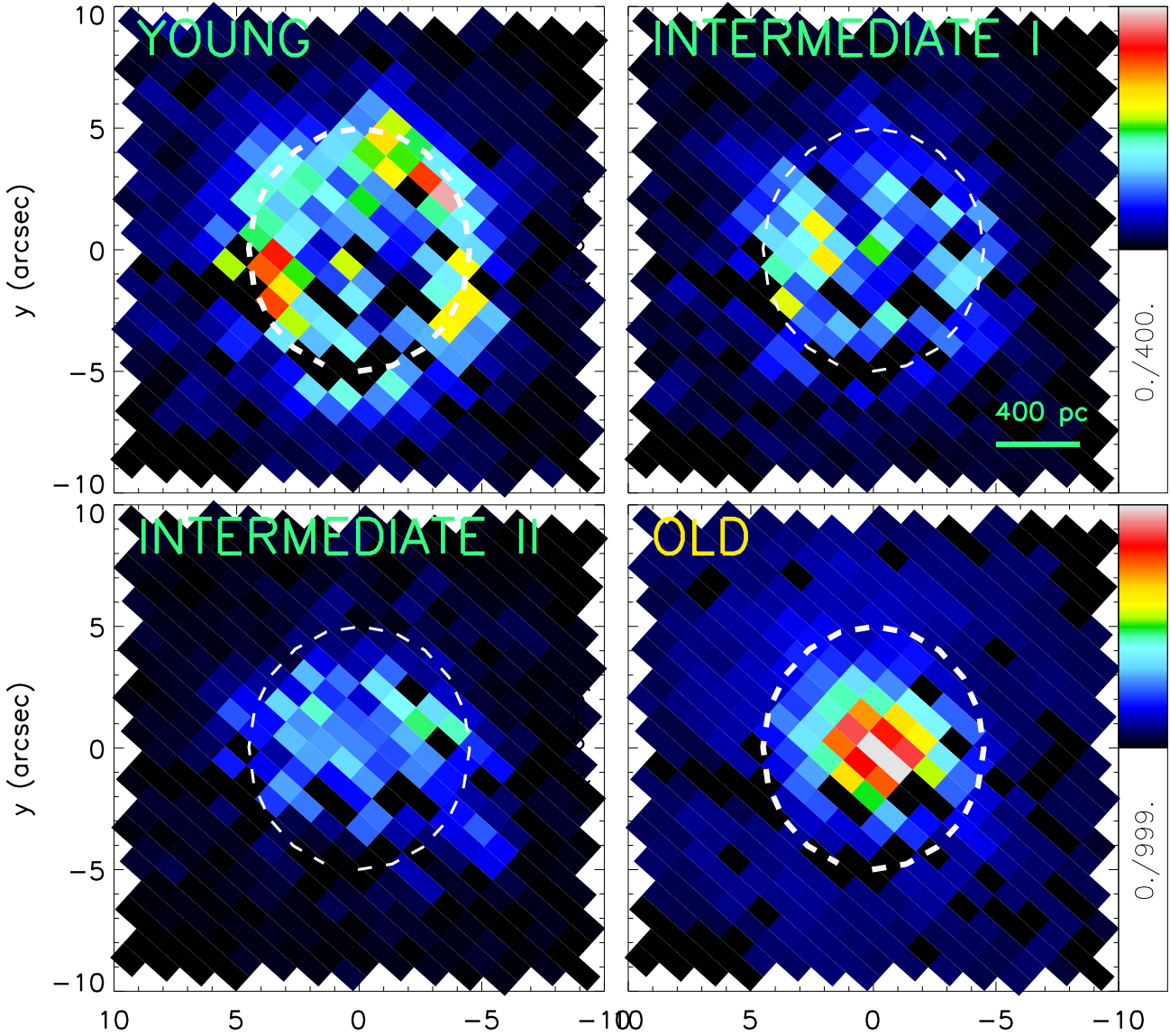}
\caption[Age distinguished stellar mass maps]{The stellar `mass' maps  in [a.u.] of mass by age bin obtained by fitting the \Sauron\ IFU dataset with a combination of 4 SSP spectra.  The maps are corrected for inclination and are oriented north up, east left. The approximate location (a$\sim 5\arcsec$, b$\sim 4.5\arcsec$) of the ring is indicated by the dashed ellipse. The `mass' uncertainty is within 10\%.}
\label{fig:IFU_SSP}
\end{figure*}

The star clusters discussed above (Sect. 6.1) have ages of several hundred Myr and are visible due to their concentrated light. How do the star clusters relate to the underlying diffuse stellar populations in the ring? Does the ring have a different star formation history from the disk, or does the disk contain similar stellar populations? With the \Sauron\  IFU dataset these questions can be addressed. 

Similar to the work by \citet{2009A&A...500.1287S}, we fit a combination of `young' and `old' single stellar population (SSP) spectra to each Vonoroi cell\footnote{At the ring positions the size of a Vonoroi cell is equal to the origin spaxel.} of the \Sauron\ mosaic using pPXF and GANDALF with similar settings as in Sect. \ref{sec:gandalf}. The goal is to split the light at each position into its `young' and `old' stellar population fractions. The assumption here is that the stellar content observed in each cell can be approximated by a finite, small, combination of SSPs. Due to the continuous evolution of stars over time, this assumption can usually be made, as long as the combination is chosen to remain sensitive to the full possible age range ($0-13$\,Gyr). SSP spectra spanning $0-13$\,Gyr were extracted from the MILES database\footnote{http://miles.iac.es} \citep{Sanchez+06,2011A&A...532A..95F} with a spectral resolution of $\sigma$ = 60\,km/s and spectral sampling of 1.15\AA{}/pixel. In order to avoid an age-metallicity degeneracy, only solar metallicity SSP models were selected. The original fit of the \Sauron\ data, discussed in Section \ref{sec:gandalf} was more extensive; a range of metallicities was used there. It showed that solar metallicity is the dominant metallicity and thus a valid simplification here. Then, using mock observations, generated from the MILES SSP spectra together with added noise, the set of SSP spectra that still reasonably span the full possible age range could be restricted to 4 SSP spectra. Their age identifiers in the MILES database are 70.8Myr, 158.5Myr, 398.1Myr and 3.2Gyr. From here on, they will be designated `young', `intermediate I', `intermediate II' and `old'. These 4 spectra will obviously not give the same age refinement as the full set, but our main goal, as stated, is to separate `young' and 'old' stellar light at each spaxel/Vonoroi cell. Each Vonoroi cell in the \Sauron\ mosaic was fit with a combination of these 4 SSPs.

The spectral fitting routine pPXF gives a (normalized) number weighted output of the SSPs used to fit the spectra at each position. We used a `bootstrap'-like method to obtain error estimates on the output generated by fitting the 4 SSPs to each spaxel/Vonoroi cell. From the fit of the data with a full set of SSP models, a measure for the fitting residual was obtained. This `residual', which is approximately 3\% of the flux, was used to introduce random noise of this magnitude into the spectra. The fitting was iterated 30 times, each time introducing new random noise. Again based on mock data, the uncertainty of the number weighted fractional contribution by each of the 4 SSPs is estimated to be within 10\% after 30 iterations. We notice from the mock data that if a large `old' population is present, the fitting underestimates the `young' fraction by several percent. The opposite is not true. The results therefore remain sensitive to older populations also in regions dominated by light from `young' stellar populations. Further, the intermediate fractions are slightly under-predicted in all cases, to the advantage of the `young' and 'old' fractions. The fitting routine attempts to minimize the number of SSPs necessary where possible. The normalized number weighted output is transformed into (again normalized) luminosity fractions, with the known luminosity of each SSP spectrum. The luminosity fractions can also be transformed into normalized stellar mass fractions, by use of the M/L ratio tabulated for each SSP spectrum. Finally, we multiply the normalized mass fractions with the stellar intensity at each cell (Fig. \ref{fig:SAURON}, top-left panel) to obtain a common baseline for all spaxels. This can be interpreted as an `absolute' mass as function of spatial position for each age bin. The resulting maps for the 4 age bins are shown in Fig. \ref{fig:IFU_SSP}.

As can be seen from Fig. \ref{fig:IFU_SSP} (results have been corrected for inclination), the ring is prominent in the `young' bin, comprising about 30-40\% of the mass at its galactocentric distance. The contact-points between the large scale bar induced gas lanes and the ring stand out at the north and south positions in the ring. The mass of `young' stars is highest at those positions. In the `old' bin the opposite picture is observed; almost no `old' stellar population is clearly associated with the ring, the `old' stars are distributed in a typical bulge-like profile. Significant mass in `intermediate' age stellar populations is found at the inner edge of the ring.

Fig. \ref{fig:IFU_SSP} shows that the ring is not circular, but slightly ellipsoidal, when corrected for inclination. Ellipticity is very common for the entire population of circumnuclear rings \citep[e.g.][]{1993AJ....105.1344B,Comeron2009}.

\section{Discussion}
\subsection{Star formation signatures in the ring}\label{sect:ringwedges}

\begin{figure*}
\includegraphics[width=17cm]{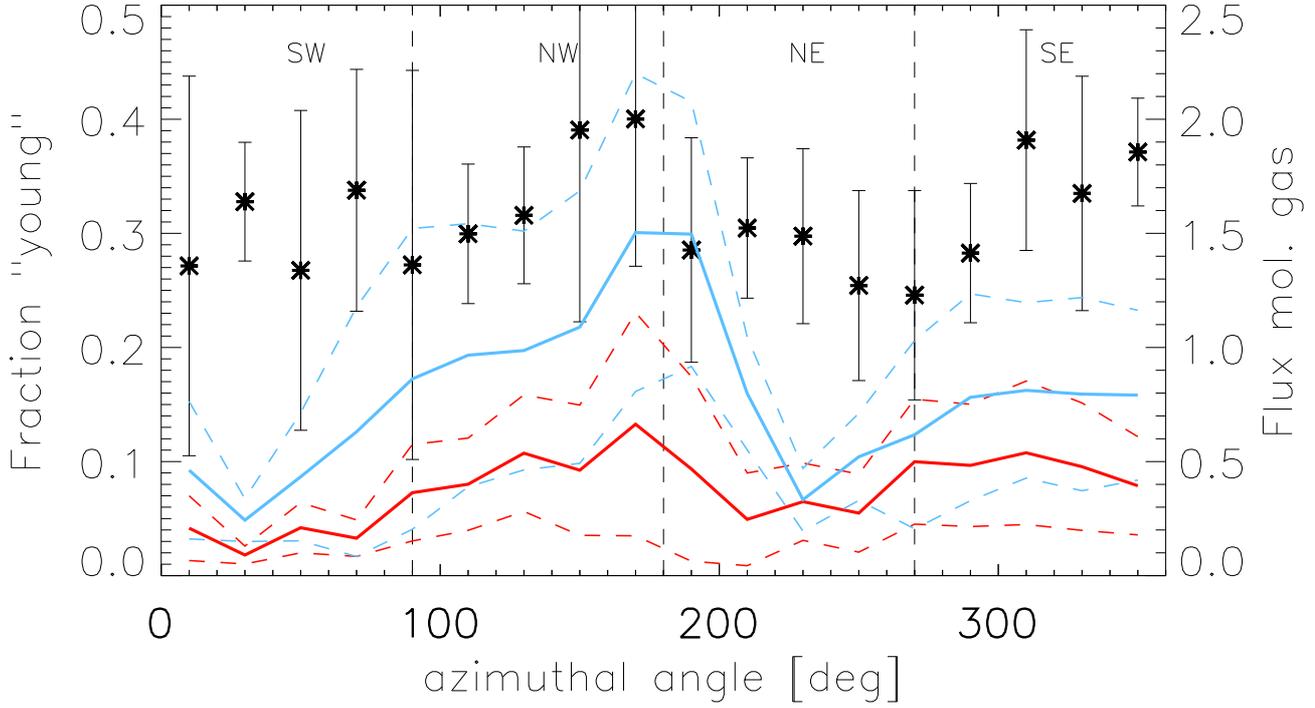}
\caption{Azimuthal mass fraction of young stars (asterisk), combined with the azimuthal intensity distribution of CO(2-1) (red line, dashed $\pm$1\,standard deviation) and HCN(1-0) emission (blue line, dashed $\pm$1\,standard deviation). The angle 0 corresponds to the south, angle increases counter-clockwise.}
\label{fig:spatial}
\end{figure*}

By combining the results from the star cluster analysis and the underlying stellar `field' population through the IFU fitting, the stellar content in and around the circumnuclear ring has been fully constrainted. What does it say about the manner in which star formation proceeds in this circumnuclear ring? Is there an azimuthal age gradient, as in the `pearls-on-a-string' scenario proposed by \citet{Boker2008}, or has star formation resulted in a homogeneous (`popcorn') stellar age distribution \citep{Elmegreen1994}?

The recent star formation episodes are concentrated in the ring region (see Fig. \ref{fig:IFU_SSP}). This is in accordance with our opening statement that the circumnuclear ring is the gas reservoir for the circumnuclear region. The `old' stellar bin in Fig. \ref{fig:IFU_SSP} mainly shows a bulge-like profile, with no additional mass at the ring radius itself. The IFU fitting shows a significant `young' mass fraction in the ring itself, between 30\% and 40\%. The average age of the star clusters in the ring is young/intermediate; 200-300\,Myr. This discrepancy in age (range) between the two methods is partly due to the IFU method, which slightly under-predicts the intermediate stellar fraction in favor of the `young' fraction. However, the IFU fitting also covers the full ring and is thus equally sensitive to star formation which did not lead to the formation of $\geq10^4M_{\odot}$ star clusters (the lightest clusters in our sample). The ring could be currently in a state where star formation leads to $\leq10^4M_{\odot}$ clusters. Possibly, the current gas inflow rate is lower than $\geq$100\,Myr ago when the detected star clusters were formed.

Both the star cluster and IFU results show that the distribution of the `young' stars is not uniform throughout the ring. The relation between the contact points and star formation is the main observable for the `pearls-on-a-string' scenario for star formation. The locations with the highest mass in `young' stars are close to the gas inflow points, the contact points, which is a point in favor of the `pearls-on-a-string' scenario.

In Fig. \ref{fig:spatial} we further plot the intensity distributions of the molecular gas tracers CO(2-1) and HCN(1-0) as a function of azimuthal angle in the ring. The azimuthal distribution of the `young' stellar mass fraction (in percentage of the total mass at each position, based on the IFU results) is also plotted there. The highest concentrations of molecular gas are close to the contact points in the north ($\phi \sim$ 180\degr) and to a lesser extent in the south ($\phi \sim$ 350\degr). The variations are especially pronounced in the HCN distribution. The fraction of the total mass in `young' stars by comparison shows much less variation.

The highest concentrations of `young' stars are found in the NW and SE segments of this figure. The wave pattern of the large scale stellar bar is slower (28.9\,km/s/kpc pattern speed) than the stellar and gas velocity ($\sim$\,150\,km/s) at the circumnuclear ring radius (580\,pc). This means that the overdensity points will move counter-clockwise in the inertial frame of the galaxy (galaxy rotation is clockwise). If stars form preferentially at or near the overdensity points, the stars and gas will move faster and out of the overdensity region. The fraction of `young' stellar mass must thus be higher in front of the overdensity region (in Fig. \ref{fig:spatial} 'in front of' equals at {\it lower} azimuthal angle). The offset between the highest fractions of `young' stars and the peak in HCN emission in the north is $\sim$ 20\degr . Approximately 1.5\,Myr would be needed to travel 20\degr\, at the ring radius. 

A half (two-fold symmetry) orbit of the bar wave pattern in the ring takes approximately 100\,Myr,  and stars that have formed at those locations will make 5 full orbits around the ring before the pattern rotation is complete. Hence, any age signature will be severely smeared within that 100\,Myr pattern time, and erase the `pearls-on-a-string' scenario in favor of the `popcorn' one. The small indications of young fractions we find are thus all we can detect.

\subsection{Age of the circumnuclear ring}
The stellar age content of the ring can assist in the determination of the age of the ring as a dynamical structure. Taking together the stellar ages, gas inflow rate and dynamical times allows both a minimum and maximum age to be constrained.

The lack of extra mass in the ring in the old stellar bin in Fig. \ref{fig:IFU_SSP} suggests that the ring cannot be older than $\sim$3\,Gyr. The average age of the star clusters is 200-300\,Myr with some older (up to 1\,Gyr). \citet{Tessel} already inferred a minimum age of the large scale stellar bar and thus the ring of 1.1\,Gyr, based on mass inflow and total molecular gas mass (after helium correction). 

\begin{figure}
\resizebox{\hsize}{!}{\includegraphics{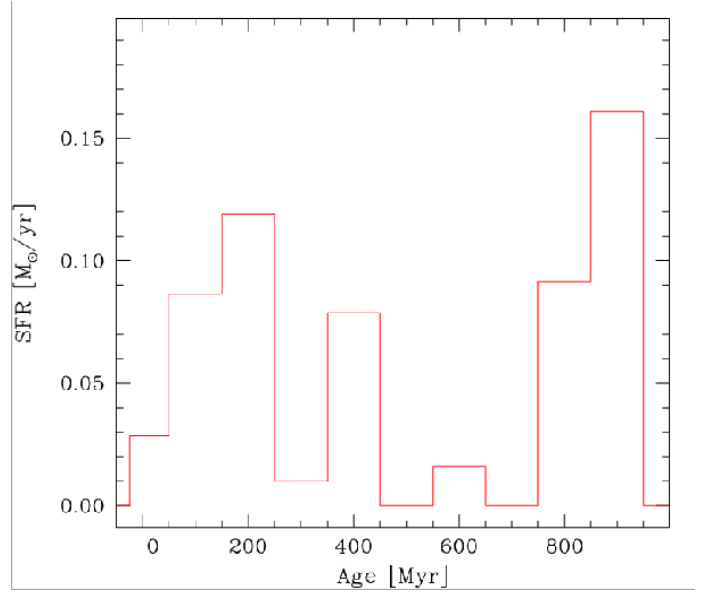}}
\caption[Star formation rate history in the circumnuclear ring]{Star formation rate history in the circumnuclear ring. The masses of the star clusters are grouped in 100\,Myr age bins and divided by the width of the age bin. No correction has been applied for cluster disruption, which would increase the SFR at older ages.}
\label{fig:SFR}
\end{figure}

The inflow rate determined in \citet{Tessel} was 2.0$M_{\odot}$/yr. The total mass in the star clusters we found is $\sim$ 6$\times10^7$M$_{\odot}$. A star (cluster) formation rate history, based on the star cluster results, is given in Fig. \ref{fig:SFR}. The highest SFR was reached 800\,Myr ago, after which the ring was quiescent for 300\,Myr. Star formation has started again in the last 400\,Myr. Over 1\,Gyr (the oldest measured star cluster age), the average star (cluster) formation rate was $\sim$ 0.06$M_{\odot}$/yr. A star formation efficiency of a few percent is entirely plausible. That implies that the current inflow rate is a good estimate of the average inflow rate over the last 1\,Gyr, even though the star formation rate was variable, and that the previously inferred minimum age of 1.1\,Gyr, based on that inflow rate, is still plausible. If the star formation efficiency is higher, a lower inflow rate would suffice and the minimum age would be higher.

The fraction of angular momentum lost by gas per orbit, dL/L, is 0.2-0.4 \citep{Haan2009}, which means gas needs at least 3-5 orbits to settle on the ring, which equals $\sim$200\,Myr (orbit times are larger at larger radii). 

The oldest likely age for the ring is 3\,Gyr, given the lack of a significant number of `old' age stars from the analysis in Sect. \ref{sect:IFUstar}. If the ring were 3\,Gyr, that implies that the ring was quietly accumulating gas for up to 2\,Gyr before star formation in the ring commenced. Assuming a constant mass inflow of 2.0$M_{\odot}$/yr and a ring of 600\,pc (5\arcsec) radius and 100\,pc (1\arcsec) width, that would imply an absurdly high molecular gas surface density of 11,000\,M$_{\odot}$/pc$^2$ at the onset of star formation. That is much higher than the average 100\,M$_{\odot}$/pc$^2$ value of most giant molecular clouds \citep[e.g.][]{2008AJ....136.2846B}. To reach an average surface density of 100\,M$_{\odot}$/pc$^2$ would thus only take 19\,Myr.

The current age of the circumnuclear ring is therefore most likely between 1.0\,Gyr and 1.5\,Gyr.

\section{Conclusions}\label{sect:conclusions}
In this work the properties of the stellar populations of the circumnuclear ring in NGC\,6951 were investigated. The goals were to estimate the age of the circumnuclear ring itself, to determine the age of the stellar population(s) present in the circumnuclear ring, and to determine whether the `popcorn' or 'pearls-on-a-string' scenario holds true for the nature of star formation in the circumnuclear ring. The main results can be summarized as follows:

The ring is most likely between 1.0\,Gyr and 1.5\,Gyr old. It has a stellar population distinct from the rest of the circumnuclear region that was formed over that time frame. The maximum possible age for the ring is 3\,Gyr. However, it is very unlikely that the ring has been in place for that long without forming stars for its first 1.5\,Gyrs.

Analysis of the general stellar content of the ring, by means of the \Sauron\ mosaic, shows that 30-40\% of the stellar mass in the ring is from `young' ($<$100\,Myr) stars. The largest fractions are found near the contact points that link the ring with the large-scale stellar bar induced gas lanes.

The majority of the star clusters in the circumnuclear ring have intermediate ages. The youngest star clusters ($\sim$100-200\,Myr), when the results are averaged over 45\degr\, intervals, are in the north and south, close to the contact points were gas flows onto the ring. Such a connection is indicative of the `pearls-on-a-string' scenario for star formation in circumnuclear rings. However, the dynamical time of the ring is short; it takes only 24\,Myr for a star cluster to complete one rotation of the ring, while the age gradient can only be renewed every 100\,Myr. Therefore we do not find definite evidence of the `pearls-on-a-string' scenario in the circumnuclear ring of NGC\,6951.

\begin{acknowledgements}
The authors thank the referee, Jes{\'u}s Falc{\'o}n-Barroso, for helpful comments that improved this paper. The authors also thank M. Krips for sending a version of her HCN map. We acknowledge the ING for supporting \Sauron\, and executing the \Oasis\, observations in service mode. Some of the data presented in this paper were obtained from the Multimission Archive at the Space Telescope Science Institute (MAST). STScI is operated by the Association of Universities for Research in Astronomy, Inc., under NASA contract NAS5-26555. Support for MAST for non-\textit{HST} data is provided by the NASA Office of Space Science via grant NNX09AF08G and by other grants and contracts. TvdL wishes to thank A. Adamo for valuable help and discussion on $\chi^2$ fitting. TvdL acknowledges DFG-funding (grant SCHI 536/2-3). RMcD is supported by the Gemini Observatory, which is operated by the Association of Universities for Research in Astronomy, Inc., on behalf of the international Gemini partnership of Argentina, Australia, Brazil, Canada, Chile, the United Kingdom, and the United States of America.
\end{acknowledgements}

\bibliographystyle{aa}
\bibliography{20285.bib}

\clearpage \onecolumn
\begin{landscape}
\begin{longtable}{l c c c c c c c c c c c c c c c c}
\caption{\label{tab:fluxiraf} Star clusters identified with IRAF/DAOPHOT and SExtractor} \\
\hline\hline
\multicolumn{1}{c}{ID} & $\Delta\alpha$ [''] & $\Delta\delta$ [''] & M$_{330}$ & $\sigma_{330}$ & M$_{547}$ & $\sigma_{547}$ & M$_{606}$ & $\sigma_{606}$ & M$_{658}$ & $\sigma_{658}$ & M$_{814}$ & $\sigma_{814}$ & M$_{110}$ & $\sigma_{110}$ & M$_{160}$ & $\sigma_{160}$ \\ \hline
\endfirsthead
\multicolumn{17}{c}%
{\tablename\ \thetable{} -- continued from previous page} \\
\hline \hline \multicolumn{1}{c}{ID} & $\Delta\alpha$ [''] & $\Delta\delta$ [''] & M$_{330}$ & $\sigma_{330}$ & M$_{547}$ & $\sigma_{547}$ & M$_{606}$ & $\sigma_{606}$ & M$_{658}$ & $\sigma_{658}$ & M$_{814}$ & $\sigma_{814}$ & M$_{110}$ & $\sigma_{110}$ & M$_{160}$ & $\sigma_{160}$ \\ \hline
\endhead
\multicolumn{17}{c}{{-- continued on next page}} \\ 
\endfoot

\hline
\endlastfoot
1 &  -0.5 &   4.4 &  -10.2 &  0.03  &  -10.4 &    0.06 &   -10.3 &    0.04 &   -10.5 &    0.48 &   -10.0 &    0.11  &   -9.2 &    0.41 &    -8.6  &   0.40 \\
  &  -0.5  &   4.4  &  -10.3 &    0.03 &   -10.6 &    0.07 &   -10.5 &    0.05 &   -11.0 &    0.46 &   -10.3 &    0.12 &    -9.8 &    0.38 &    -9.5 &    0.34 \\
2 & -0.8  &   4.2 &   -10.0 &    0.03 &    -9.4 &    0.10 &    -9.0 &    0.08  &   -9.7 &    0.68  &   -8.5 &    0.23 &    -8.3 &    0.62 &    -7.4 &    0.71 \\
3 & -2.8 &    4.1 &    -5.6 &    1.49 &   -11.4 &    0.04  &    \nodata & \nodata &    -6.8 &    3.28 &    -5.0 &    2.20 &     \nodata &    \nodata &     \nodata &    \nodata \\
4 & -1.3   &  4.1 &   -11.0 &    0.03  &  -10.7 &    0.07 &   -10.5  &   0.08 &   -11.0 &    0.47 &   -10.0 &    0.16 &     \nodata &    \nodata &     \nodata &    \nodata \\
5  &  -1.3  &   4.0 &   -11.4 &    0.02 &   -10.7 &    0.05 &   -10.7 &    0.04 &   -11.3 &    0.33  &  -10.1 &    0.11 &    -8.5 &    0.56 &    -7.9 &    0.55 \\
  &  -1.3  &   4.0 &   -11.1 &    0.02 &   -10.9 &    0.06 &   -10.8 &    0.05 &   -11.5  &   0.37 &   -10.4 &    0.12 &     \nodata &    \nodata  &    \nodata &    \nodata \\
6 &   -1.0 &    3.9  &  -10.6 &    0.02 &   -10.0 &    0.07 &    -9.9&     0.05 &   -11.7 &    0.28 &    -8.9  &   0.19 &    -7.5 &    0.90  &   -6.6 &    1.03 \\
 &   -1.0  &   3.9 &   -10.7 &    0.03 &   -10.3 &    0.09 &   -10.4 &    0.07 &   -12.0 &    0.28 &    -9.5 &    0.20 &     \nodata &    \nodata  &    \nodata &    \nodata \\
7 &   -1.6 &    3.9 &   -10.0 &    0.03 &   -10.0  &   0.07 &   -10.1 &    0.05 &   -10.8 &    0.42 &   -10.0 &    0.11  &   -8.4 &    0.61 &    -7.1 &    0.80 \\
  &   -1.5 &    3.9 &   -10.5 &    0.05 &   -10.5 &    0.09 &   -10.5 &    0.08 &   -11.0 &    0.46 &   -10.1 &    0.16 &    -9.8 &    0.41 &    -9.0 &    0.54 \\
8 &   -0.4 &    3.7  &  -10.5 &    0.03 &   -11.0 &    0.04 &   -11.0 &    0.03 &   -11.0 &    0.39 &   -10.9  &   0.08 &   -10.2 &    0.26 &    -9.8 &    0.23 \\
  &   -0.4 &    3.7 &   -10.5 &    0.03 &   -11.1 &    0.05  &  -11.0 &    0.04 &   -11.2 &    0.42  &  -10.8  &   0.09 &   -10.2 &    0.33 &    -9.9 &    0.32 \\
9 &   -1.6 &    3.6 &   -10.5 &    0.02  &   -9.4 &    0.09 &    -9.3 &    0.07 &   -10.6 &    0.47 &    -8.4 &    0.24 &    -7.4 &    0.95 &    -6.2 &    1.21 \\
10 &   -3.5 &    3.6 &   -10.7 &    0.02 &   -10.1 &    0.07 &   -10.1 &    0.04 &   -10.5 &    0.48 &   -10.4  &   0.10  &   -9.8 &    0.31 &    -9.1 &    0.32 \\ 
  &   -3.5 &    3.6  &  -10.5 &    0.02 &   -10.2  &   0.08  &  -10.3 &    0.05 &   -11.1 &    0.44 &   -10.5 &    0.11&    -10.0 &    0.33  &   -9.6 &    0.33 \\
11*  &  -2.1  &   3.4 &     \nodata&    \nodata &     \nodata &   \nodata  &    \nodata&    \nodata  &    \nodata&    \nodata &     \nodata &   \nodata &   -10.3 &    0.36 &   -10.0  &   0.32 \\
12 &   -2.0 &    3.3 &   -10.5 &    0.03 &   -10.6  &   0.05 & -10.5  &   0.04 & -10.2  &   0.56 & -10.2  &   0.10  & -9.7  &   0.32  & -9.4  &   0.27 \\
  & -2.0 &   3.3 & -10.4  & 0.03 & -10.6  & 0.06 & -10.6  & 0.06 & -10.8  & 0.50 & -10.4  & 0.12 & -10.0  & 0.36  & -9.7  & 0.34 \\
13 & -1.1 &    3.2  & -8.8  & 0.12  & -9.9  & 0.08 & -10.0  & 0.05 & -10.3  & 0.52 & -10.1  & 0.11  & -9.6  & 0.34  & -9.1  & 0.32 \\ 
14* & -3.6 &    3.1   & \nodata & \nodata   & \nodata & \nodata   & \nodata & \nodata   & \nodata & \nodata   & \nodata & \nodata & -10.5  & 0.33 & -10.3  & 0.27 \\
15 & -2.1&     2.9 & -10.7  & 0.02 & -10.5  & 0.05 & -10.6  & 0.04 & -10.8  & 0.42 & -10.3  & 0.10  & -9.7  & 0.33  & -9.1  & 0.31 \\
 & -2.1 &    2.9 & -10.6  & 0.02 & -10.5  & 0.07 & -10.5  & 0.05 & -10.7  & 0.53 & -10.3  & 0.12  & -9.6  & 0.45  & -9.3  & 0.43 \\
16* &    1.9 &    2.8   & \nodata & \nodata   & \nodata & \nodata   & \nodata & \nodata   & \nodata & \nodata   & \nodata & \nodata  & -9.1  & 0.65  & -8.6  & 0.64 \\
17 &    1.6 &    2.6  & -9.8  & 0.05  & -9.7  & 0.09  & -9.8  & 0.05 & -10.5  & 0.47  & -8.9  & 0.19  & -7.5  & 0.92  & -7.1  & 0.80 \\
18  &   1.0 &    2.4  & -9.8  & 0.05  & -9.7  & 0.08  & -9.4  & 0.07 & -10.5  & 0.49  & -9.0  & 0.18  & -8.4  & 0.59  & -8.2  & 0.49 \\
19 & -4.3  &   2.4 & -10.4  & 0.02 & -10.2  & 0.07 & -10.0  & 0.05  & -9.9  & 0.63  & -9.7  & 0.13  & -9.3  & 0.38  & -9.2  & 0.31 \\
  & -4.3 &    2.4 & -10.7  & 0.02 & -10.5  & 0.06 & -10.4  & 0.05 & -10.5  & 0.56 & -10.3  & 0.12 & -10.0  & 0.34  & -9.9  & 0.28 \\
20 &     1.3 &    2.3  & -9.3  & 0.06 & -11.2  & 0.04  & -8.8  & 0.09  & -9.9  & 0.62  & -7.5  & 0.36  & -7.0  & 1.14  & -4.7 &  2.42 \\
 &    1.3 &    2.3  & -9.2  & 0.08 & -11.4  & 0.05  & -8.8  & 0.26 & -10.4  & 0.60  & -7.7  & 0.70    & \nodata  & \nodata    & \nodata  & \nodata \\
21 & -4.3 &    2.2 & -10.0  & 0.03  & -9.7  & 0.08  & -9.8  & 0.05 & -10.0  & 0.60  & -9.1  & 0.17  & -8.8  & 0.49  & -8.3  & 0.47 \\
  & -4.3 &    2.2 & -10.4  & 0.03 & -10.3  & 0.07 & -10.3  & 0.05 & -10.6  & 0.54 & -10.1  & 0.13    & \nodata  & \nodata    & \nodata  & \nodata \\
22 & -4.1 &    2.2 & -11.6  & 0.01 & -11.4  & 0.04 & -11.3  & 0.03 & -11.4  & 0.32 & -10.9  & 0.07 & -10.2  & 0.26  & -9.7  & 0.24 \\
  & -4.1 &    2.2 & -11.7  & 0.01 & -11.7  & 0.04 & -11.5  & 0.03 & -11.6  & 0.34 & -11.2  & 0.08 & -10.5  & 0.26 & -10.3  & 0.24 \\
23 &    2.5 &    1.8  & -9.6  & 0.05  & -9.5  & 0.09  & -9.6  & 0.06  & -9.4  & 0.81  & -9.4  & 0.15  & -9.0  & 0.46  & -8.3  & 0.47 \\
24  &   1.6  &   1.8  & -6.7  & 0.29 & -13.3  & 0.01  & -7.0  & 0.21  & -7.0 &  2.37  & -5.4  & 0.97  & -3.3 &  6.31  & -4.9 &  2.23 \\
   &   1.6 &    1.8  & -7.0  & 0.48 & -13.7  & 0.01  & -7.2  & 1.32  & -8.5 &  1.46    & \nodata & \nodata & \nodata  & \nodata    & \nodata  & \nodata \\
25  &   2.5  &   1.6 & -10.3  & 0.03 & -10.4  & 0.06 & -10.3  & 0.04 & -10.5  & 0.47 & -10.0  & 0.11  & -9.9  & 0.30  & -9.2  & 0.30 \\
  &    2.6  &   1.6  & -9.9  & 0.04 & -10.2  & 0.08 & -10.1  & 0.07 & -10.4  & 0.60 & -10.0  & 0.14  & -9.7  & 0.40  & -9.4  & 0.38 \\
26  &   2.5 &    1.1  & -9.4  & 0.06  & -9.9  & 0.07  & -9.8  & 0.05 & -10.2  & 0.56  & -9.7  & 0.13  & -9.2  & 0.41  & -8.5  & 0.42 \\
27 & -3.6 &    1.0  & -8.8 & 10.00  & -9.4  & 0.11  & -9.4  & 0.09 & -10.6  & 0.55  & -8.9  & 0.23    & \nodata  & \nodata    & \nodata  & \nodata \\
28 & -3.8  & 0.5  & -9.9  & 0.04 & -10.4  & 0.06 & -10.3  & 0.04 & -10.9  & 0.40 & -10.2  & 0.10  & -9.2  & 0.41  & -8.5  & 0.42 \\
  & -3.8  & 0.5  & -9.7  & 0.05 & -10.5  & 0.07 & -10.5  & 0.05 & -11.0  & 0.46 & -10.3  & 0.12    & \nodata  & \nodata    & \nodata  & \nodata \\
29 & -3.7  & 0.3 & -11.1  & 0.02 & -10.9  & 0.05 & -10.7  & 0.03 & -10.8  & 0.42 & -10.4  & 0.10  & -8.8  & 0.49  & -7.6  & 0.62 \\
  & -3.7  & 0.3 & -11.2  & 0.02 & -11.1  & 0.05 & -11.0  & 0.04 & -11.3  & 0.39 & -10.6  & 0.10  & -9.6  & 0.40  & -9.0  & 0.45 \\
30 &  3.2 & -0.3 & -10.7  & 0.02 & -10.6  & 0.05 & -10.7  & 0.03 & -12.4  & 0.21 & -10.1  & 0.11  & -8.9  & 0.48  & -8.0  & 0.54 \\
   & 3.2 & -0.3 & -10.2  & 0.03 & -10.1  & 0.08 & -10.2  & 0.06 & -11.9  & 0.29  & -9.8  & 0.15  & -9.1  & 0.52  & -8.3  & 0.68 \\
31  & 0.2 & -0.3  & -9.6  & 0.06 & -10.9  & 0.05 & -11.5  & 0.02 & -13.2  & 0.14 & -11.3  & 0.06 & -11.4  & 0.15 & -11.2  & 0.12 \\
32 & -5.5 & -0.4  & -9.3  & 0.04 & -10.4  & 0.06 & -10.4  & 0.04 & -10.4  & 0.50 & -10.2  & 0.10  & -8.9  & 0.46  & -7.9  & 0.55 \\
  & -5.5 & -0.4  & -9.5  & 0.04 & -10.7  & 0.06 & -10.7  & 0.04 & -10.8  & 0.50 & -10.5  & 0.10  & -9.4  & 0.43  & -8.7  & 0.48 \\
33 & -4.9 & -0.5  & -9.2  & 0.06 & -10.2  & 0.08 & -10.2  & 0.05 & -10.2  & 0.66  & -9.9  & 0.14  & -9.0  & 0.54  & -8.5  & 0.54 \\
34 &    3.4 & -1.4 & -10.0  & 0.03 & -10.9  & 0.05 & -11.1  & 0.03 & -12.0  & 0.24 & -10.9  & 0.08 & -10.0  & 0.29  & -8.7  & 0.37 \\
    &   3.4 & -1.4  & -9.9  & 0.04 & -11.0  & 0.05 & -11.1  & 0.04 & -12.1  & 0.27 & -11.0  & 0.09  & -9.9  & 0.36  & -8.9  & 0.51 \\
35* & -4.4 & -1.7   & \nodata & \nodata   & \nodata & \nodata   & \nodata & \nodata   & \nodata & \nodata   & \nodata & \nodata  & -9.5  & 0.57  & -9.2  & 0.52 \\
36* & -1.8 & -1.8   & \nodata & \nodata   & \nodata & \nodata   & \nodata & \nodata   & \nodata & \nodata   & \nodata & \nodata  & -9.9  & 0.46  & -9.5  & 0.42 \\
37* &    3.9 & -2.1   & \nodata & \nodata   & \nodata & \nodata   & \nodata & \nodata   & \nodata & \nodata   & \nodata & \nodata  & -9.7  & 0.54  & -9.2  & 0.50 \\
38 &    4.3 & -2.1 & -11.0  & 0.02 & -10.8  & 0.05 & -10.7  & 0.03 & -10.7  & 0.44 & -10.3  & 0.10  & -9.4  & 0.38  & -9.0  & 0.33 \\
    &    4.3 & -2.1 & -11.0  & 0.02 & -11.0  & 0.05 & -10.9  & 0.04 & -11.1  & 0.44 & -10.6  & 0.10 & -10.1  & 0.32  & -9.9  & 0.28 \\
39 & -1.5 & -2.5  & -9.6  & 0.05  & -9.8  & 0.08  & -9.7  & 0.06 & -10.1  & 0.57 & -10.2  & 0.10 & -10.0  & 0.28 & -10.0  & 0.21 \\
  & -1.6 & -2.5  & -9.9  & 0.05 & -10.4  & 0.08 & -10.4  & 0.07 & -10.6  & 0.56 & -10.4  & 0.12 & -10.1  & 0.33 & -10.1  & 0.27 \\
40 &    4.2 & -2.5  & -9.9  & 0.05 & -10.5  & 0.06 & -10.5  & 0.04 & -10.4  & 0.50 & -10.2  & 0.11  & -9.6  & 0.34  & -9.1  & 0.32 \\
41 & -1.8 & -2.5  & -9.8  & 0.03  & -9.0  & 0.11  & -9.1  & 0.07  & -9.3  & 0.83  & -9.1  & 0.17  & -9.3  & 0.40  & -9.0  & 0.33 \\
   & -1.8 & -2.5 & -10.2  & 0.04 & -10.3  & 0.09 & -10.2  & 0.08 & -10.2  & 0.65 & -10.0  & 0.15  & -9.9  & 0.36  & -9.8  & 0.32 \\
42 & -1.5 & -2.6  & -9.6  & 0.04  & -9.5  & 0.09  & -9.3  & 0.07 & -10.0  & 0.60  & -8.7  & 0.21  & -8.8  & 0.48  & -8.4  & 0.43 \\
43  &   2.8 & -2.7 & -10.3  & 0.03 & -10.4  & 0.06 & -10.3  & 0.04 & -10.2  & 0.56  & -9.9  & 0.12  & -8.6  & 0.54  & -7.8  & 0.59 \\
44 &    1.4 & -2.9  & -9.5  & 0.06  & -9.6  & 0.14  & -9.4  & 0.14  & -9.4  & 0.94  & -9.3  & 0.22  & -8.8  & 0.65  & -8.4  & 0.64 \\
45 &    3.3 & -3.0 & -10.1  & 0.03  & -9.9  & 0.07  & -9.8  & 0.05 & -10.1  & 0.59  & -9.6  & 0.13  & -8.9  & 0.47  & -8.5  & 0.43 \\
   &    3.3 & -3.0 & -10.1  & 0.03 & -10.1  & 0.08 & -10.1  & 0.06 & -10.6  & 0.55  & -9.8  & 0.15    & \nodata  & \nodata    & \nodata  & \nodata \\
46 & -1.8 & -3.1  & -9.4  & 0.08  & -9.9  & 0.08  & -9.9  & 0.05  & -9.8  & 0.66  & -9.6  & 0.13  & -8.6  & 0.54  & -8.2  & 0.49 \\
47 &  2.8 & -3.2  & -9.9  & 0.03  & -9.2  & 0.10  & -9.3  & 0.07  & -9.8  & 0.65  & -8.8  & 0.20  & -8.2  & 0.64  & -7.7  & 0.62 \\
    &   2.8 & -3.2 & -10.5  & 0.03 & -10.2  & 0.09 & -10.1  & 0.09 & -11.0  & 0.46  & -9.6  & 0.20    & \nodata  & \nodata    & \nodata  & \nodata \\
48 & -0.6 & -3.2 & -10.1  & 0.03 & -10.3  & 0.09 & -10.1  & 0.09 & -10.2  & 0.66  & -9.6  & 0.20  & -9.2  & 0.57  & -9.0  & 0.52 \\
  & -0.7 & -3.2 & -10.4  & 0.03 & -10.6  & 0.05 & -10.4  & 0.04 & -10.5  & 0.47 & -10.1  & 0.11  & -9.3  & 0.38  & -8.7  & 0.38 \\
49  & 0.3 & -3.6 & -10.1  & 0.04  & -9.7  & 0.08  & -9.6  & 0.06  & -9.8  & 0.67  & -9.5  & 0.14  & -9.6  & 0.35  & -9.1  & 0.31 \\
\*  & 0.2 & -3.5   & \nodata & \nodata   & \nodata & \nodata   & \nodata & \nodata   & \nodata & \nodata   & \nodata & \nodata & -10.2  & 0.37  & -9.1  & 0.57 \\
50*  &   1.6 & -3.5   & \nodata & \nodata   & \nodata & \nodata   & \nodata & \nodata   & \nodata & \nodata   & \nodata & \nodata & -10.5  & 0.35 & -10.1  & 0.32 \\
51 &    2.7 & -3.6 & -11.6  & 0.01 & -11.2  & 0.04 & -11.0  & 0.03 & -11.2  & 0.35 & -10.6  & 0.09  & -9.3  & 0.39  & -8.1  & 0.51 \\
   &    2.7 & -3.6 & -11.4  & 0.01 & -11.2  & 0.05 & -11.1  & 0.04 & -11.4  & 0.38 & -10.5  & 0.11  & -9.9  & 0.37  & -8.9  & 0.57 \\
52  & 2.3 & -3.7  & -9.7  & 0.07 & -10.1  & 0.12 & -10.1  & 0.12 & -10.9  & 0.49  & -9.7  & 0.21    & \nodata  & \nodata    & \nodata  & \nodata \\
53  & 0.2 & -3.8  & -9.3  & 0.06 & -10.0  & 0.07  & -9.9  & 0.05 & -10.0  & 0.60 & -10.1  & 0.11  & -9.9  & 0.30  & -9.6  & 0.26 \\
   & 0.1 & -3.8   & \nodata & \nodata   & \nodata & \nodata   & \nodata & \nodata   & \nodata & \nodata   & \nodata & \nodata & -11.4  & 0.20 & -10.9  & 0.19 \\
54 &    1.1 & -3.8  & -9.6  & 0.06  & -9.9  & 0.08  & -9.8  & 0.05 & -10.1  & 0.57  & -9.9  & 0.12  & -9.1  & 0.43  & -8.6  & 0.40 \\
55  &   1.1 & -4.1  & -9.6  & 0.07  & -9.9  & 0.08  & -9.9  & 0.05 & -11.1  & 0.37  & -9.5  & 0.15  & -8.8  & 0.49  & -8.2  & 0.49 \\
\end{longtable} %*
\textbf{Notes:} Candidate star clusters detected with IRAF/DAOPHOT. Col (1) star cluster candidate ID; Col. (2)-(3) projected offset in arcseconds from the photometric nucleus (RA 20$^{h}$37$^{m}$14.123$^{s}$, Dec 66\degr06\arcmin20.09\arcsec); Col. (4)-(15) absolute magnitude (m-M=31.9), M$_{x}$, and error in magnitude, $\sigma_{x}$, in the F330W, F547M, F606W, F814W, F110W and F160W band \textit{HST} images. The candidates with a star in column 1 could not be fitted with an acceptable star cluster model in (see \S\ref{sect:stellarages}).

\end{landscape}

\clearpage \onecolumn
\begin{longtable}{l c c c c}
\caption{Star cluster ages, extinctions and masses}\label{tab:clusterfitresult} \\
\hline\hline
\multicolumn{1}{c}{ID} & Age [Myr] & E(B-V) & log(Mass/M$_{\odot}$) & $\chi^2$/d.f. \\ \hline
\endfirsthead
\multicolumn{5}{c}%
{\tablename\ \thetable{} -- continued from previous page} \\
\hline \hline \multicolumn{1}{c}{ID} & Age [Myr] & E(B-V) & log(Mass/M$_{\odot}$) & $\chi^2$/d.f. \\ \hline
\endhead
\multicolumn{5}{c}{{-- Continued on next page}} \\ 
\endfoot

\hline
\endlastfoot
  1    &   265    &    0.44    &     5.9    &  0.56 \\
  2    &    36    &    0.02    &     4.6    &  1.13   \\
  3    &   964    &    0.00    &     6.4    &  5.01   \\
  4    &    88    &    0.38    &     5.6    &  0.60   \\
  5    &    78    &    0.35    &     5.6    &  3.90   \\
  6    &    60    &    1.15    &     5.5    & 14.08   \\ 
  7    &   184    &    0.80    &     6.0    &  2.50   \\
  8    &   440    &    0.62    &     6.4    &  1.25   \\
  9    &     6    &    0.31    &     4.0    &  2.87   \\
10    &    75    &    1.14    &     5.8    &  3.71   \\
12    &   220    &    0.72    &     6.1    &  1.11   \\
13    &   998    &    0.93    &     6.4    &  1.11   \\
15    &   130    &    0.90    &     6.0    &  0.72   \\
17    &   196    &    0.50    &     5.6    &  4.20   \\
18    &   130    &    0.23    &     5.2    &  2.02   \\
19    &   122    &    0.40    &     5.6    &  1.54   \\
20    &   864    &    0.00    &     6.1    &174.29   \\
21    &   130    &    0.56    &     5.6    &  1.65   \\
22    &   140    &    0.41    &     6.1    &  0.37   \\
23    &   148    &    0.83    &     5.6    &  0.11   \\
24    &   265    &    0.00    &     6.8    &340.26   \\
25    &   234    &    0.65    &     5.9    &  1.52   \\
26    &   420    &    0.43    &     5.8    &  1.24   \\
27    &     4    &    1.26    &     4.6    &  0.84   \\
28    &   700    &    0.49    &     6.2    &  2.29   \\
29    &   135    &    0.30    &     5.8    &  3.29   \\
30    &   134    &    1.00    &     6.0    & 21.18   \\
31    &   828    &    0.78    &     6.8    &125.41   \\
32    &   998    &    0.75    &     6.5    &  6.87   \\
33    &   998    &    0.44    &     6.2    &  1.68   \\
34    &   996    &    0.90    &     6.8    &  8.48   \\
38    &   145    &    0.51    &     5.9    &  0.54   \\
39    &   420    &    1.04    &     6.4    &  4.04   \\
40    &   490    &    0.60    &     6.2    &  0.12   \\
41    &   100    &    1.04    &     5.5    &  2.45   \\
42    &   136    &    0.31    &     5.2    &  1.32   \\
43    &   216    &    0.42    &     5.8    &  1.15   \\
44    &   190    &    0.69    &     5.6    &  0.37   \\
45    &   162    &    0.50    &     5.6    &  0.49   \\
46    &   450    &    0.36    &     5.8    &  0.52   \\
47    &    64    &    0.64    &     5.3    &  0.53   \\
48    &   241    &    0.21    &     5.7    &  0.48   \\
49    &   122    &    0.43    &     5.4    &  5.49   \\
51    &   100    &    0.18    &     5.7    &  1.97   \\
52    &   336    &    0.72    &     6.0    &  0.68   \\
53    &   836    &    0.62    &     6.2    &  5.88   \\
54    &   248    &    1.04    &     6.0    &  2.38   \\
55    &   294    &    0.58    &     5.8    &  2.85   \\
\end{longtable}
\textbf{Notes:} Results of the $\chi^2$ fitting of age (column 2), color excess (column 3) and mass (column 4) of the observed star clusters. Only those candidates which were successfully fitted are listed. The star cluster ID in column 1 is equal to that in Table \ref{tab:fluxiraf}.\\

\end{document}